# Does the body's mass depend upon its speed?
## (Discussion via exchange of letters)

## M. Ya. Amusia[1, 2]

[1] *Racah Institute of physics, The Hebrew University, 91904 Jerusalem, Israel*
[2] *Ioffe Physico-Technical Institute,194021 St. Petersburg, Russia*

**Abstract**

The presented letters covers an almost year-long discussion of the author and a Very Qualified scientist (VQS)[1] about the dependence of mass upon speed if relativistic corrections are taken into account.

VQS believes that since mass is a scalar, it cannot depend upon speed and has to be the same in all inertial coordinate frames. In his view, the very idea of speed-dependence of the mass of a particle or a body is incorrect and misleading. As such, the notion of speed dependence of a particle mass has to be eliminated from textbooks on physics and from teaching of this subject.

The author claims that this notion has the right to exist, is easily understandable and convenient for most of the students, both non-physicists and even physicist. His view is that it is nothing wrong in expressions like "particle mass increases with the growth of speed". His view upon the debate is that both approaches are equally correct. His view is that the dilemma "depends or not depends" is a matter of taste and convenience, not of scientific importance and unavoidable scientific rigor.

Upon request of VQS, M. Amusia has commented the manuscript of academician L. B. Okun's resent book «ABC of physics». The comments are included into this letters exchange.

**Аннотация**

Приведенные ниже письма охватывают длившееся почти год, путём переписки, обсуждение автором и Очень квалифицированным учёным, (ОКУ)[2] зависимости массы от скорости при учёте релятивистских поправок.

ОКУ считает, что поскольку масса является скаляром, она не может зависеть от скорости и должна быть одинакова во всех инерциальных системах отсчёта. По его мнению, сама идея зависимости массы частицы или тела от скорости неверна, не нужна и лишь вводит в заблуждение. Поэтому, понятие зависимости массы частицы должно быть исключено из учебников по физике и из преподавания этого предмета.

М. Я. Амусья утверждает, что это понятие имеет право на жизнь, наглядно и легко объясняемо, а потому удобно для большинства студентов, как не-физиков, так и физиков. Он считает, что ничего плохого, а тем более - неверного, в выражении "масса частиц увеличивается с ростом её скорости" нет. Его точка зрения состоит в том, что оба подхода одинаково достоверны и имеют право на жизнь. Он считает, что разница "зависит или не зависит" - дело вкуса и удобства, а не научной значимости и научной точности.

---

[1] Who preferred to remain anonymous.
[2] Который пожелал остаться анонимным.

По просьбе ОКУ, М. Я. Амусья прокомментировал рукопись недавней книги академика Льва Борисовича Окуня «АВС физики". Комментарии включены в этот обмен письмами.

# Переписка М. Я. Амусья с Очень квалифицированным учёным о зависимости массы от скорости в специальной теории относительности (СТО)

Я решил опубликовать эти совершенно не личные, а чисто научные письма в их первоначальном виде, лишь с небольшими комментариями, делающими дискуссию более понятной по возможности широкому кругу читателей. Мне кажется, что живая дискуссия с человеком, придерживающимся существенно иной точки зрения, читателю гораздо интереснее безответной критики опубликованных работ.

В последний момент мой корреспондент предпочёл не открывать своё инкогнито. Думаю, что конкретное имя придало бы переписке лишь дополнительный интерес, а также позволило бы напрямую общаться со обоими, а не только с одним, автором.

*Самое страшное это пустяк, поднятый до уровня общемировой проблемы*
А. Зиновьев[3]

Начата письмом М. Я. Амусья 31.03.11 (Далее письмо обозначается А1)

Глубокоуважаемый Леон Болеславович![4]
20.03.11 я написал Вам письмо, в котором отмечал, что:

«**В₁**. С заметным опозданием познакомился со статьёй Л. Б. Окуня "Понятие массы" в УФН[5]. Сознавая пользу аккуратных терминов, хочу обратить внимание и на цену, которая платится за уточнение. Так, "Вейники[6]" (надеюсь, Вы помните имя этого борца с энтропией?) получают возможность авторитетнейшей ссылки, способствующей распространению их трудов. Примером, на мой взгляд, служит статья теплотехника И. Эткина о его "Энергодинамике". Она появилась в ряде не имеющих прямого отношения к физике журналов и сайтов, хотя вместо этого могла бы стать предметом рассмотрения Комиссии по борьбе с лженаукой РАН под председательством Э. П. Круглякова[7].

**В₂**. Так же с опозданием, увы, теперь неисправимым, обнаружил статью Е. Л. Фейнберга о "некинематичности" лоренцева сокращения длины и замедления времени. Ни секунды не сомневаюсь, что ЕЛ хотел сказать нечто правильное о реальности сокращения и замедления. Но звучит для постороннего уха странно, что процесс ускорения системы координат, происходивший, скажем, 567 лет сему назад, сказывается как-то на длинах и временах процессов здесь и сейчас. Не мудрено, что и этим воспользуется немало антинаучников. Их ссылки на признанные авторитеты помогают, увы, распространению бредней. Вроде статьи В. Миркина "Бог не играет в кости с физиками", и без того сильно облегчённой наличием Интернета.

---

Хотел бы услышать Ваше мнение по этому аспекту проблемы уточнения формулировок».

Вы в ответ поинтересовались моим мнением о том, зависит ли масса от скорости и просили уточнить, что именно мне больше всего не нравится в упомянутой заметке Е.Л. Фейнберга. Я сообщил, что мне кажутся вводящими в недоумение разговоры о реальных изменениях, происходящих при лоренцовом сокращении. Вводит, на мой взгляд, в заблуждение и разговор о прошлом ускорении системы отсчёта, движущейся рядом с лежащим на столе металлическим, например, прутком. По-моему, ровно ничего в нём не произошло, как и в нашем инерциальном трамвае, который начал двигаться, и набрал рассматриваемую скорость сто лет назад. Разумеется, можно сказать, что таких длинных рельс в природе нет, но всё это удаляет от рассмотрения и описания давно известного и хорошо понятого эффекта СТО[8].

Тем временем, я прочитал рекомендованные Вами статьи Л. Б. Окуня не только в УФН 158, 511-530 (1989), но и в УФН 170, 1366-1371 (2000), 178, 541-555, 663-663 (2008)[9]. Мне было интересно, но определённо, признаюсь, удивило внимание к вопросу, который считал до прочтения этих статей, да и продолжаю считать сейчас, чисто терминологическим. Но Ваше внимание к нему понуждает меня ответить на заданный вопрос о моём отношении к зависимости массы от скорости, и, одновременно, прокомментировать ряд утверждений из упомянутых выше статей Л. Б. Окуня. Итак, по порядку:

1. В лекциях и популярных объяснениях пользуюсь понятием зависимости массы от скорости и не сталкиваюсь при этом с трудностями.
2. Говорю студентам, что в самой релятивистской теории вполне уместно ограничиться лишь массой покоя, пользуясь релятивистскими выражениями для энергии и импульса и законами их сохранения.
3. Всегда отмечаю, что автоматическая подстановка вместо массы покоя $m_0$ выражения $m = m_0 / \sqrt{(1 - v^2 / c^2)}$ [10] любые формулы классической механики, без разбора, не делает их автоматически релятивистскими, как не делает подобная подстановка в уравнение Шредингера его релятивистским.
4. Нередко отмечаю, что и эффективную массу, которая естественно возникает в теории многих тел, скажем, однородного электронного газа нельзя подставлять вместо электронной массы в формулу для плазменной частоты. Такая подстановка грубо ошибочна.
5. Широко пользуюсь техникой диаграмм Фейнмана в их нерелятивистской версии.
6. Коллеги по еврейскому университету, которым я задал Ваш вопрос[11], преподающие СТО, возраста 40-60 и старше, пользуются одновременно двумя массами и двумя подходами, считая разницу между ними чисто терминологической.
7. Ни я, ни они не видят "безобразия" формулы для энергии $E = mc^2 = m_0 c^2 / \sqrt{(1 - v^2 / c^2)} \equiv m_0 \gamma c^2$. Лично я работаю в атомной системе единиц и мне система единиц с $c = 1$ не удобна. Вообще, выбор специфической системы единиц всё-таки подходит для профессионалов в области. Не думаю, что

---

изложение релятивистской квантовой механики удобно начинать с выражения $p = 1$, которое, хотя и правильно, но очень трудно объяснять.

8. Система единиц $\hbar = c = 1$ кажется мне не очень удобной для обычных студентов, не аспирантов, поскольку затрудняют переход к классике и нерелятивизму. Не могу согласиться с утверждением Л. Б. Окуня, что "$c$ в качестве единицы скорости навязывается самой природой, когда мы хотим рассматривать фундаментальные процессы в ней". Всё-таки всё это дело соглашения.

9. Упоминаемое Л. Б. Окунем как непоследовательность или объясняемое иными причинами сосуществование у Фейнмана в его книгах, Ландау (в книгах с Е. М. Лифшицем или Ю. Б. Румером), да и у самого Эйнштейна обоих подходов к массе, считаю проявлением их отношения к вопросу как чисто терминологическому.

10. Фраза Ландау об его с Румером книге как о месте, где "два жулика уговаривают третьего, что за гривенник он может понять, что такое теория относительности" отнюдь не отражает, на мой взгляд, отношения Ландау к общей книге. Ведь третьего - читателя, можно полагать невеждой, но не жуликом. Скорее, речь идёт, если такое высказывание и имело место, об общих трудностях популяризации науки.

11. Вот и Л. Б. Окунь, говоря об энергии атома, делит энергию входящего в него электрона на кинетическую и потенциальную, и вводит скорость атомного электрона, определённым значением которой он, как известно, не обладает.

12. Не очень нагляден и образ античастицы, как частицы, движущейся вспять во времени, предложенный, но не развитый, кстати, впервые Я. И. Френкелем.

13. Мне кажется, что многократное повторение в двух последних статьях одного подхода как "единственно правильного" не будет способствовать его распространению.

14. Не подходят, на мой взгляд, для дискуссии в УФН слова Л. Б. Окуня "сорняк массы, зависящей от скорости", или сопоставление использованного Э. Неизвестным в памятнике Хрущёву белого и чёрного мраморов с отношением к двум видам массы. Не годится и однозначное отнесение "массы покоя" к словам-паразитам. Если же говорить о лингвистике, то любой язык содержит "слова-паразиты", без которых языковые пуритане хотят обойтись. У них в итоге получается нежизнеспособная канцелярщина.

15. Не вижу, что, как пишет Л. Б. Окунь, в 20 веке шла "борьба двух концепций массы". Это сильное преувеличение.

16. Задевающе звучат слова Л. Б. Окуня "*Пришла пора прекратить обманывать всё новые и новые поколения школьников и студентов* (курсив мой - МА), внушая им, что возрастание массы с увеличением скорости - это экспериментальный факт". Думаю, что нормальные преподаватели и научные работники - не невежды, и никого не обманывают. Также задевает фраза Л. Б. Окуня: "*Мне представляется, что недопустимо выдавать зависимость массы от скорости за экспериментальный факт, скрывая от студента* (курсив мой - МА), что она является интерпретационным фактоидом".

17. Отмечу, что если сказать школьнику или студенту, начинающему изучать физику, будто согласно СТО энергия нарастает с увеличением скорости, то это вызовет у него недоумение: "Причём тут СТО? Ведь рост энергии с увеличением скорости следует и из известной с этак 7го класса школы формулы для кинетической энергии $E_{kin} = mv^2/2$.

18. Вы отмечаете, что Дж. А. Уилер много сделал для установления "понятия лоренц-инвариантной массы". Соответствующей ссылки у Л. Б. Окуня не нашёл.

19. Считаю запись основного соотношения СТО в виде $m^2 = e^2 - p^2$ неудобной, несмотря на упоминание Л. Б. Окунем того факта, что рекламный щит на Рублёвском шоссе около Москвы содержит формулу Эйнштейна с малым $e$. Предпочитаю сохранить $e$ как обозначение заряда электрона, а $E$-энергии.

20. С. У. Хокинг не просто, как пишет Л. Б. Окунь, популяризатор теории относительности. Ссылка на него в контексте обсуждаемого вопроса уместна не только из-за упоминания на суперобложке русского издания его книги того факта, что она переведена на 40 языков. Существенно, что книга пропагандирует формулу $E = mc^2$. Но ведь это не криминал, а свидетельство определённой точки зрения, хотя от точки зрения Л. Б. Окуня и отличающейся.

21. Не возьму в толк, в противоречии с тем, что Вы пишите, как представление о релятивистской массе может помешать пониманию механизма передачи всей энергии быстро движущегося электрона покоящемуся. Иное дело, как ведёт себя с ростом энергии вероятность такого процесса.

22. На мой взгляд, это очень хорошо, что в учебники и научные статьи не ввели обсуждаемые Л. Б. Окунем понятия "инерции" или "инера", "тяжести" или "грава". От этого путаницы было бы невпроворот. Побольше, чем от "конгруентных" треугольников и квадратов там, где в них нет ни нужды, ни квалифицированных кадров для разъяснения.

23. Не замечал ничего в поддержку утверждения Л. Б. Окуня "Забавно, что обычно те, кто сравнивает механику Ньютона с астрологией, верят, что масса зависит от скорости". Обратное определённо неверно - вижу это у себя и многих коллег.

24. Думаю, что в изложении многих областей физики, включая и такую важную, как СТО, уместно руководствоваться известным стишком

> *Мой милый, тебе эту песню дарю.*
> *Рассчитывай силы свои*
> *И если сказать ты не можешь "хрю-хрю",*
> *Визжи, не стесняясь, "и-и!".*

Относя себя к "визгливым поросятам", я и-и - каю, оказываясь, при этом, отнюдь не в самой плохой компании - старых, и не только, специалистов в обсуждаемой области. Конечно, мода и воззрения меняются,

> *и то, что было ясность мудреца,*
> *Потомкам станет бредом сумасшедших...*

Но замечательно написал Дюрренматт: "Нет ничего сомнительнее, чем вера, которая подавляет сомнение".

25. Заочно полемизируя с Л. Б. Окунем, я, наверное, провоцирую резкую отповедь. Типа "ты, Зин, на грубость нарываешься". Но свои точки зрения скрывать не привык. Да и возраст виляние не поощряет. Кроме того, ещё раз обращаю Ваше внимание на то, с чего начал -- с расползания анти- и псевдо-науки и того, как там используются научные работы и критика признанно великих, о чём говорил и что имел ввиду в своём исходном письме.

С глубочайшим уважением, искренне Ваш Мирон Янкелевич Амусья.

Письмо М. Я. Амусья 06.04.11 (А2)

## О зависимости массы от скорости в Специальной Теории Относительности СТО (2)

Глубокоуважаемый Леон Болеславович!

Прошу прощения за задержку с ответом. Но я прочитал статью Л. Б. Окуня в Архиве[12] и обмен его писем с Л. П. Питаевским. Для полноты ответа повторю свою точку зрения, изложенную, возможно, и несколько другими словами, в письме от 31.03.11.

1. Вопрос, об использовании или неиспользовании понятия массы, зависящей от скорости – вопрос удобства обозначения, и не касается существа дела.

2. Этот вопрос абсолютно не следует увязывать с глубокой и фундаментальной проблемой природы массы, и объяснения специфического значения массы той или иной «элементарнрой» частицы.

3. Вопрос о природе массы для своего решения может потребовать, в принципе, и отказа от принципа релятивистской или Лоренц-инвариантности теории.

4. Несомненно, что можно изложить специальную теорию относительности, не вводя понятия массы, зависящей от скорости.

5. Именно так я понимаю слова Л. Д. Ландау, упомянутые Л. П. Питаевским, о том, что «масса, зависящая от скорости в Курсе не нужна». Действительно, Питаевский пишет: «I directly asked Dau[13] about absence in the Curse the "velocity depended mass" He answered that it is completely useless notion». Мне представляется очевидным, что речь идёт о Курсе теоретической физики Ландау и Лифшица, который рассчитан отнюдь не на начинающих или просто любителей предмета.

6. Вы ссылаетесь на Володю Грибова, который разделял взгляды Л. Б. Окуня, но советовал ему «не бороться с этим неверным уравнением[14], «отговаривал его вести борьбу против понятия массы, зависящей от скорости». Я вижу в этом свидетельство того, что Грибов не находил здесь важной научной проблемы, но видел только некий терминологический спор. Исхожу при этом из того, что сам Грибов был в критических замечаниях прям и нелицеприятен. Например, вступил в полемику с Дираком на институтском семинаре о знаке перед массой в уравнении Дирака (и оказался прав), публично критиковал неравенства Белла как ненужные для обоснования справедливости квантовой механики и т.п. Он резко выступал против могущественных людей, от которых зависели и его научные регалии, если считал себя правым. Думаю, из-за отношенческих проблем не был избран сразу в академики, когда ему не хватило для этого всего одного голоса.

7. Сам с Грибовым о массе не говорил, но слышал его доклады и лекции, где фигурировало и качественное описание следствий релятивистских эффектов.

8. Поскольку формулы $E = m_0 \gamma c^2$ и $\vec{p} = m_0 \gamma \vec{v}$ верны, естественно ввести обозначение $m = m_0 \gamma$, приводящее к более компактным соотношениям $E = mc^2$ и $\vec{p} = m\vec{v}$. Этим специалисты при кратком описании СТО и её достижений пользовались десятилетия, а многие пользуются и сейчас и будут пользоваться в будущем. Разумеется, надо отмечать опасность автоматической, без знания СТО, замены массы в *любой* нерелятивистской формуле для полного учёта релятивизма.

---

[12] Имеется в виду электронная библиотека Корнелльского университета США - arXiv.org

[13] Принятое среди друзей имя Л. Д. Ландау

[14] Имеется в виду уравнение $E = mc^2$, которое вовсе не неверно, тождественно совпадая с $E = m_0 \gamma c^2$ при $m \equiv m_0 \gamma$.

9.  Л. Д. Ландау сам вводил эффективную массу в теории конденсираванного состояния на основе аналогии со свободным движением. При этом понятно, что гидродинамика низких температур не получится просто путём замены всюду в теории разреженного газа реальной массы на эффективную.

10. Посмотрел статью Зураба Силагадзе. Меня удивили его опасения, что из-за заголовка "Brainwashed by Newton?" его статью не примут в Архив. Такого я не слыхивал. Хотя заглавие и не кажется мне удачным. Заодно в сети посмотрел ещё несколько его статей. Конкретное содержание данной работы вряд ли кого-то дополнительно убедит в пользе того или иного обозначения. Что касается идеи преподавания исходя из логики предмета, а не истории его развития – эта мысль очевидна, и ею руководствуются очень многие. Но иногда логика совпадает с историей. Так, вводя нерелятивистские диаграммы Фейнмана, я следую весьма близко его исходным работам. Разумеется, мало кто преподаёт электричество на основе движения лягушачьей лапки.

11. Отмечу, что историческое не всегда означает «отсталое». Например, пока ещё не привилась попытка объяснять структуру нормальных ядер при средних и низких энергиях их взаимодействия на языке кварков и глюонов, а физика молекул опирается в большой мере на атомы и межатомные силы, а не на смеси ядер и электронов.

12. Можно основы физики изложить и на ста страницах, основываясь на двух константах – постоянной Планка и скорости света в пустоте. Но трудно будет читателю понять соотношение $\approx \hbar c / 137.04$ вместо привычного элементарного (возможно, пока) заряда наблюдаемой элементарной частицы электрона $e$. Позволю себе здесь привести анекдот. Старый еврей говорит: «Верю, что Иисус накормил народ двумя хлебами. Но вот стали ли они сыты – это большой вопрос».

13. «Ликвидация» термина «масса, зависящая от скорости» не кажется мне признаком даже терминологической чистоты. Ведь мы нередко работаем, избирая определённую систему координат – движущуюся, что подчёркивается явной зависимостью $m = m(v)$. Вполне сознавая независимость электродинамического результата от калибровки, я обычно выбираю Кулоновскую, в которой атомные процессы рассматривать гораздо удобнее, чем в любой другой. Вы же едва ли ею пользуетесь.

14. Не вижу, каким образом приведённая Вами «релятивистская» кружка[15] принижаем статус науки в обществе. Скорее, наоборот. Выбрали научный орнамент – и хорошо. Вряд ли стоит полемики надпись на кружке, юбке, майке и т.п., если они не оскорбительны для посторонних, не имеют погромного или расистского содержания. А вот обвинение профессоров, не разделяющих точку зрения Л. Б. Окуня, которой в популярном изложении не следовали ни Фейнман, ни Ландау, ни многие другие, в том, что они «вводят в заблуждение школьных преподавателей и их учеников», кажется мне несправедливым.

15. Вы пишите: «Утверждение о том, что масса тела должна меняться с его скоростью, *заведомо неверное* (курсив мой – МА)». Поскольку считаю вопрос по сути терминологическим, оснований для столь сильного утверждения не вижу. Приводимый Вами пример с почти одинаковым удержанием магнитным полем ускорителя в ЦЕРНе[16] протонов и электронов при их огромной разности масс покоя просто объясняется качественным замечанием о том, что масса обоих частиц почти одинакова, поскольку их скорость очень близка к скорости света.

---

[15] Имеется в виду фаянсовая кружка с нанесёнными на ней формулами и утверждениями, обычно относимыми к теории относительности.
[16] Европейский центр ядерных исследований.

16. Ещё раз обращаю Ваше внимание на поступь антинаучников – видя, как терминологический спор крупнейшим физиком Л. Б. Окунем поднимается до принципиального, они говорят: «Смотрите, сами физики запутываются в трёх соснах. А мы им даём «единые и универсальные теории всего», и они не принимают этого. Косные они и просто путаются в простейших вещах». Два примера подобных работ я Вам привёл. Но их не два, а, увы, легион. Это подрывает доверие к науке в обществе куда больше, чем надпись на кружке. Подрывает это доверие и участие учёных в сомнительных кампаниях, типа холодного термояда, глобального потепления и т.п.

17. Позволю несколько отойти в сторону, и отметить иную интерпретацию фразы из книги В. Паули, приводимой Е. Л. Фейнбергом в его статье о некинематичности лоренцева сокращения и замедления времени. Паули писал: «Сокращение масштаба является не простым, а … крайне сложным процессом» Когда все законы, «определяющие строение электрона», станут известными, «теория будет в состоянии дать атомистическое объяснение». На мой взгляд, здесь нет речи о прошлом ускорении одной из систем отсчёта – иначе, почему упоминается лишь электрон, а не другие составляющие тела. Думаю, что Паули имел в виду некое чудо. Оно состоит в том, что все динамические законы, определяющие строение электрона и его взаимодействие, оказываются Лоренц – инвариантны. А ведь могло, в принципе, быть иначе, и Лоренц- инвариантность, при переходе к какому-либо новому взаимодействию, оказывалась бы несправедливой, подобно тому, как электромагнитное взаимодействие не укладывалось в рамки галилеевой инвариантности.

18. Отвлекаясь в сторону от нашего сюжета, отмечу, что в своей статье УФН 1989 г. Л. Б. Окунь упоминает письмо секретаря Шведской академии наук Эйнштейну. Секретарь писал, что Эйнштейн награждён Нобелевской премией за открытие теории фотоэффекта, но не учитывая ценности, которая будет признана за теориями относительности и гравитации, «после того, как они в будущем будут подтверждены». Вот описание того, за что дали премию: The Nobel Prize in Physics 1921 was awarded to Albert Einstein *"for his services to Theoretical Physics, and especially for his discovery of the law of the photoelectric effect"*. А вот официальное мнение Председателя Нобелевского комитета Сванте Аррениуса, взятое из его речи при награждении Эйнштейна в декабре 1922 г.: «There is probably no physicist living today whose name has become so widely known as that of Albert Einstein. Most discussion centers on his theory of relativity. This pertains essentially to epistemology and has therefore been the subject of lively debate in philosophical circles. It will be no secret that the famous philosopher Bergson in Paris has challenged this theory, while other philosophers have acclaimed it wholeheartedly. The theory in question also has astrophysical implications which are being rigorously examined at the present time». Как видно, речь идёт о сопротивлении философов, включая Нобелевского лауреата Бергсона. Возражений против СТО в тот период среди ведущих физиков не было никаких. А вот против квантовой теории излучения как объясняющей природу света, против эйнштейновской теории фотоэффекта – были. Этому интереснейшему вопросу я посвятил лекцию, читанную в ряде стран. С 2005г. собираюсь написать на этой основе статью в УФН, да всё не соберусь.

С глубочайшим уважением, искренне Ваш Мирон Янкелевич Амусья.

**Письма: Амусья 31.03.11 -- (А1), ОКУ 07.04.11-- (О1), Амусья 30.04.11--(А2) и ОКУ 15.06.11--(О2)**

В этом разделе приведена дискусия, последовавшая за письмом М. Я Амусья **А1**. Ответное письмо ОКУ обозначено **О1**, куда Амусья вставил свои возражения, отмечаемые как А2. А затем ОКУ вставил туда же свои пояснения **О2** и соответствующие части письма **А1**, состоявшего из вступительной части **В**, и основной части, содержавшей 25 возражений, приведенных Амусья **А1₁₋₂₅**. Ответ обозначен наклонной чертой, перед которой – номер письма – ответа, а после которой – то, на что отвечается

## Вступительная часть

**О1/А1**: Глубокоуважаемый Мирон Янкелевич,

Большое спасибо за Ваше письмо от 31 марта. Мне кажется, что оно даёт прекрасную возможность обсудить поднятый Вами вопрос и приблизиться к истине. Ниже я постараюсь ответить на все Ваши утверждения (на два утверждения во вступительной части Вашего письма и на 25 пунктов основной части[17]). Я предлагаю Вам ответить на мои вопросы, и сам отвечу на Ваши. Это, может быть, единственный путь выяснения истины.

На мои реплики ко вступительной части Вы не ответили. Мои реплики ко вступительной части помечены как О1/А1$_{В1}$, О1/А1$_{В2}$ и О2/А1$_{В1}$, О2/А1$_{В2}$, чтобы отличить их от реплик в основной части.

О1/А1$_{В1}$. Вы пишете, что сознаёте пользу аккуратных терминов, но по-видимому считаете, что статья Л. Б. Окуня в УФН 1989 года, в которой он пытался разъяснить и уточнить современное понятие массы, принесла не столько пользу, сколько вред. Примером этого вреда, по Вашему мнению, является статья В. Эткина, автор которой сослался на статью Л. Б. Окуня и поддержал его утверждение, что в современной физике термин масса тела должен относиться только к такой величине, которая не зависит от скорости тела.

В письме от 21 марта я поинтересовался Вашим мнением о том, зависит ли масса от скорости. В ответ Вы сообщаете мне, что прочли ещё две статьи Л. Б. Окуня по этому вопросу от 2008 года, а также статьи Байерляйна и Хехта. (Я их читал, когда они вышли!) И сообщаете мне, что Вас удивляло и удивляет внимание к вопросу, который Вы считали и продолжаете считать чисто терминологическим. Значит ли это, что Вы считаете, что он не заслуживает внимания?

А2/О1/А1$_{В1}$. Он заслуживает умеренного внимания, главным образом, пояснения, что «говорить» можно и так, и так.

О1/А1$_{В2}$. Я полностью согласен с Вами (а Л. Б. Окунь не устаёт это подчёркивать в течение более двадцати лет), что вопрос о массе чисто терминологический. Ни я, ни Вы не сомневаемся, что теория относительности - наука правильная, и вопрос заключается в том, как излагать её и как знакомить с ней поколения наших внуков и правнуков. А здесь, я считаю, пренебрежение терминологией ведёт к недопустимому непониманию. Ведь на этой теории (и на квантовой механике) основана вся современная наука и техника. И поэтому терминология должна быть максимально точной. Если Вы согласны с этим, то почему Вы удивляетесь? Или Вы считаете, что терминология не должна быть однозначной?

---

[17] Далее они обозначены, как А1$_{В1-2}$ и А1$_{1-25}$, соответственно

А2/О1/А1$_{В2}$. Терминология, как ни старайся, совсем однозначной не будет. Причина здесь в том, что нет абсолютно однозначных словесных определений самых основных понятий физики – пространства, времени, энергии. Аккуратность в терминах нужна, но полной однозначности – не достичь просто из-за особенностей человеческого языка.

О2/А1$_{В1}$. Ответьте, пожалуйста: считаете ли Вы, что терминологический вопрос не заслуживает внимания? Это принципиальный пункт расхождения между нами. Я считаю, что терминология должна быть максимально адекватной и отточенной. Она должна быть однозначной. Недопустимо, чтобы в одних книгах по самой точной физической науке, какой является СТО, утверждалось, что масса не зависит от скорости, а в других, что масса зависит от скорости.

А2/О2/А1$_{В1}$. Терминоогия уж совсем однозначной не будет никогда, сколь её ни оттачивай. В обсуждаемом вопросе, на мой взгляд, отвергаемый Л. Б. Окунем и Вами и столь же им и Вами резко критикуемый подход, имеет полное право на жизнь.

О2/А2. Уже после того, как мы начали эту переписку, Вы сообщили мне, что принадлежите к школе В.А. Фока. Я очень уважаю труды Владимира Александровича. Я знаю, что Л. Б. Окунь потратил два десятилетия, чтобы восстановить его приоритет в вопросе о калибровочной инвариантности - см. УФН. Август 2010, стр. 871-873). Я считаю, что он был прав, когда на страницах своей книги "Теория пространства, времени и тяготения" спорил с Эйнштейном о принципе эквивалентности гравитации и ускорения в ОТО[18]. К сожалению, в этой же книге он потратил много параграфов, пытаясь пропагандировать "релятивистскую массу" и "массу покоя" в СТО. Правда, что в отличие от Вас, Фок обозначал массу покоя буквой $m$, а релятивисткую массу -- буквой $M$ (см., например, параграф 25), а обозначения $m_0$ у него не было.

Кстати, не можете ли Вы сообщить мне, в какой из работ Фока содержится замечательная фраза "Физика наука простая, надо только понимать, что какая буква обозначает"?

А2/О2/А2. Я не принадлежу к школе В. А. Фока, хотя слушал его лекции, сдавал ему экзамен, посещал его семинар и докладывал докторскую диссертацию, на которую он писал по собственной инициативе отзыв. Не разделяю Вашего сожаления по поводу использования им понятия массы, зависящей от скорости. Его книга для меня вся – просто образец того, какой должна быть научная книга. Разумеется, обозначения я у Фока не перенимал.

## Основная часть

## Ответы ОКУ и дискуссия по 25 замечаниям из письма Амусья А1 – А1$_{1-25}$.[19] Система обозначений та же, что и ранее
**Замечание А1$_1$ из письма Амусья А1**

---

[18] ОТО – общая теория относительности
[19] Система обозначений утверждение-возражение та же, что и выше.

A1$_1$.  В лекциях и популярных объяснениях пользуюсь понятием зависимости массы от скорости и не сталкиваюсь при этом с трудностями.

О1/А1$_1$. Вы пишете, что в своих лекциях и научно-популярных статьях Вы пользуетесь понятием массы, зависящей от скорости, и не сталкиваетесь при этом с трудностями. Именно поэтому обсуждение этого вопроса с Вами представляется мне особенно полезным.

Разумеется, для Вас никаких трудностей нет. Вы, как и я, изучали теорию относительности по "`Теории поля'" Ландау и Лифшица и, следовательно, понимаете её. С трудностями будут сталкиваться Ваши читатели и слушатели, которые книгу Ландау и Лифшица не читали и которые не знают, что в ней нет ни термина релятивистская масса (т.е. масса, зависящая от скорости), ни термина масса покоя. Зачем Вы прививаете им эти термины и связанную с ними систему понятий, которая была в ходу до того, как в 1941 году появилась книга Ландау и Лифшица? Какой человек или какая книга обучили Вас этим терминам и побуждают Вас быть их защитником и пропагандистом сегодня? За моей спиной стоят Ландау и Померанчук, а кто стоит за Вашей спиной? (Предыдущие фразы О1/А1$_1$ были написаны до того, как я узнал, что Вы выучили СТО по книге Фока, а не по книге Ландау и Лифшица.)

А2/О1/А1$_1$. Мне никого не нужно "за спиной". Теории относительности всех, включая Ландау и Померанчука (впервые появляющегося в нашей переписке как сторонника Вашей и Л. Б. Окуня точки зрения) обучал А. Эйнштейн. Сам Ландау был "двулик", так как ошибался, когда писал с Румером, и был прав в соавторстве с Лифшицем. "Грешили" и Фейнман с Фоком, справедливо не видя здесь ничего, кроме небольшой терминологической проблемы.

Понятие массы, зависящей от скорости, удобно и внятно представляет качественный результат СТО. Задав на недавней конференции, где был, десятку теоретиков в ранге не ниже профессора, вопрос - "Зависит ли масса от скорости", неизменно получал ответ "Согласно СТО - зависит". Не вижу революции 1941 г., в части соотношения релятивистской массы и массы покоя.

ОКУ, 10.06.12: Мне пришлось разбить Ваш ответ А2/О1/А1$_1$ на мои вопросы О1/А1$_1$ на шесть пунктов, чтобы ниже ответить на каждый из них.

А2/О1/А1$_{1.1}$ Мне никого не нужно "за спиной".

А2/О1/А1$_{1.2}$.Теории относительности всех, включая Ландау и Померанчука (впервые появляющегося в нашей переписке как сторонника Вашей точки зрения) обучал А. Эйнштейн.

А2/О1/А1$_{1.3}$. Сам Ландау был "двулик", так как ошибался, когда писал с Румером, и был прав, в соавторстве с Лифшицем. "Грешили" и Фейнман с Фоком, справедливо не видя здесь ничего, кроме небольшой терминологической проблемы.

А2/О1/А1$_{1.4}$. Понятие массы, зависящей от скорости, удобно и внятно представляет качественный результат СТО.

А2/О1/А1$_{1.5}$. Задав на недавней конференции, где был, десятку теоретиков в ранге не ниже профессора, вопрос - "Зависит ли масса от скорости", неизменно получал ответ "Согласно СТО - зависит".

А2/О1/А1$_{1.6}$. Не вижу революции 1941 г., в части соотношения релятивистской массы и массы покоя.

О2/А2/О1/А1$_{1.1}$. Вы пишете, что сдавали экзамены по теорию относительности Друкареву и Фоку. По-видимому, они и привили Вам термины "релятивистская масса" и "масса покоя" и они стоят за Вашей спиной. Но так как в книге Фока обозначения $m_0$ нет, то кто ещё стоит за Вашей спиной? Из какой книги она попала в Ваш арсенал?

А3/О2/А2/О1/А1$_{1.1}$. Не из какой. У меня всегда для массы обозначение $m$. Покой считаю естественным подчёркивать дополнительным нулём. Если есть в задаче более тяжёлая частица, то она $M$. Букв изредка нехватает – когда в задачке есть и проекция углового момента.

О2/А2/О1/А1$_{1.2}$. Я выучил теорию относительности по книге Ландау и Лифшица и слушал лекции Померанчука как раз в то время, когда Померанчук изучал статьи Фейнмана, Дайсона, Швингера по квантовой электродинамике, в которых формировалась современная терминология. В них терминов "релятивистская масса" и "масса покоя" не было, была просто масса. Померанчук умер в 1966 году, Ландау - в 1968 году. Они не были "сторонниками Л. Б. Окуня или моей точки зрения". Я был и остаюсь их учеником.

А3/О2/А2/О1/А1$_{1.2}$. Уверен, что, выучившись, Вы очень многое переосмыслили сами, так что Ваша точка зрения принадлежит Вам, не являясь слепком взглядов Ландау, Померанчука и других. Отмечу, как бы в скобках, что в статьях Фейнмана, Дайсона, Швингера по квантовой электродинамике формировалась вовсе не «современная терминология», а сама квантовая электродинамика. Соответственно, исходя из нужд этой, в значительной части тогда новой, науки, формулировалось заимствованное от предшественников. Ясно, что совать массу, зависящую от скорости в функцию Грина электрона, было бы неуместно. Они и строили ковариантную[20] формулировку.

О2/А2/О1/А1$_{1.3}$. Все люди ошибаются и "грешат". И единственный способ продвигаться к истине - это признавать свои ошибки. Утверждение о "небольшой терминологической проблеме " принадлежит Вам, а не Фейнману и Фоку. Есть ли у Вас цитаты, подтверждающие Ваше утверждение, что они считали эту проблему "небольшой"?

А3/О2/А2/О1/А1$_{1.3}$. Цитата, вообще говоря, не есть доказательство. Но сам факт сосуществования разных выражений у одного и того же авторитетного специалиста означает его нечувствительность к проблеме, которой он явно не видит. Кстати, немало неточностей, граничащих с ошибками, есть в курсе Ландау и Лифшица. В ликвидации одной из них участвовал поневоле, как ошибавшийся вместе с классиками.

О2/А2/О1/А1$_{1.4}$. Мне кажется, что Вы неправильно используете слова "качественный результат". Качественный результат может быть получен из теории (в данном случае из СТО) без вычислений. Основной качественный результат СТО заключается в том, что масса частиц не меняется при их движении, и что свободные безмассовые частицы движутся со скоростью света. Понятие массы, зависящей от скорости, возникшее до создания СТО в рамках механики Ньютона, "внятно" препятствует пониманию этого качественного результата. Понятие массы, зависящей от скорости, возникло в конце 19 века при неудачных попытках "впихнуть" электродинамику Максвелла в механику Ньютона.

---

[20] Ковариантную, т.е. одинаково выглядевшую во всех инерциальных системах отсчёта.

А3/О2/А2/О1/А1$_{1.4}$.Считаю, что абсолютно неверно утверждение, будто «основной качественный результат СТО заключается в том, что масса частиц не меняется при их движении». Подобное утверждение не вяжется с тем, что можно считать результатом революционной теории. Либо надо признать Эйнштейна недотёпой, не понявшим, что он открыл, поскольку Ваше утверждение не упомянуто им в основных (я имею в виду «К электродинамике движущихся тел», и следующую, ключевую для понимания соотношения массы и энергии, статьи) работах по СТО. Оно скорее служит, на мой взгляд, неосновательным подтверждением личной концепции в этом вопросе.

Я продолжаю, тем временем, опрос физиков, включив сюда и специалистов по теории поля и астрофизики – абсолютное большинство считает, что масса в СТО зависит от скорости.

О2/А2/О1/А1$_{1.5}$. Какие аргументы привели "десяток теоретиков в ранге не ниже профессора", утверждавших, что "согласно СТО масса зависит от скорости"? Сколько из них являются профессионалами в современной квантовой теории поля?

А3/О2/А2/О1/А1$_{1.5}$. Никаких дополнительных аргументов мне опрошенные не приводили, да я и не просил. Как у Булгакова в «Мастере и Маргарите»: «Так было».

О2/А2/О1/А1$_{1.6}$. Революции в 1941 году не было. Просто Ландау и Лифшиц показали, что в СТО есть только одна масса $m$ и что она инвариантна относительно преобразований Лоренца. Они впервые в классической теории поля обошлись без "релятивистской массы" и "массы покоя".

А3/О2/А2/О1/А1$_{1.6}$. Повторюсь: здесь ничего невозможно «показать». Очевидно, что «так можно определить», но об «показать» и речи быть не может.

**Замечание А1$_2$ из письма Амусья А1**

А1$_2$.Говорю студентам, что в самой релятивистской теории вполне уместно ограничиться лишь массой покоя, пользуясь релятивистскими выражениями для энергии и импульса и законами их сохранения.

О1/А1$_2$. Вы пишете, что студентам Вы говорите, что уместно пользоваться лишь одним понятием массы. Но тогда я не понимаю, почему Вы называете её массой покоя, а не используете термин масса, как это делают большинство профессионалов по СТО, следуя курсу Ландау и Лифшица?

А2/О1/А1$_2$. Повторюсь - подробный курс СТО можно построить и без разговора о зависимости массы от скорости. Краткий - трудно, да и не нужно. Масса, зависящая от скорости, сохраняет важное нерелятивистское соотношение $p = mv$, что удобно и наглядно. Да и не очень существенно для большинства физиков, что делают профессионалы по СТО: эта теория давно уже достояние всей физики. Нам ведь не нужен для преподавания законов Ньютона некий особый специалист именно по этим законам.

O2/A2/O1/A1$_2$. Ниже разбиваю A2/O1/A1$_2$ на три пункта:

A2/O1/A1$_{2.1}$. Повторюсь - подробный курс СТО можно построить и без разговора о зависимости массы от скорости.

A2/O1/A1$_{2.2}$. Краткий - трудно, да и не нужно.

A2/O1/A1$_{2.3}$. Масса, зависящая от скорости, сохраняет важное нерелятивистское соотношение для импульса $p = mv$, что удобно и наглядно.

O2/A2/O1/A1$_{2.1}$. Подробные курсы СТО в "Теории поля" Ландау и Лифшица, "Квантовой электродинамике" Ахиезера и Берестецкого, "Введении в теорию квантованных полей" Боголюбова и Ширкова и "Введение в квантовую теорию поля" Пескина и Шрёдера построены без разговора о массе, зависящей от скорости. К сожалению, подробный курс СТО в книге Фока "Теория пространства, времени и тяготения", который Вы изучили студентом и сдали Фоку, весь построен "на разговоре о массе M, зависящей от скорости".

A3/O2/A2/O1/A1$_{2.1}$. Не вижу, в применение к Фоку, оснований для сожаления. Много раз повторял, что перечисленные авторы стояли не перед драматическим выбором, а перед неким предпочтением. Стоило бы обратить внимание, что именно Фок, человек весьма математического склада ума, сторонних чётких определений, что понудило его назвать ОТО «Теорией пространства, времени и тяготения», предпочёл массу, зависящую от скорости.

O2/A2/O1/A1$_{2.2}$. Объяснить, что масса не зависит от скорости, очень легко на нескольких страницах. Л. Б. Окунь и я делали это неоднократно в научно-популярных текстах за последние 20 -30 лет. Что Вас не устраивает в этих текстах? А делать это не просто нужно, а необходимо, чтобы донести до читателей очень важную мысль, что механика Эйнштейна содержит в себе механику Ньютона как свой предельный случай при малых скоростях. И что при больших скоростях уравнения Ньютона просто неверны.

A3/02/A2/O1/A1$_{2.2}$. *Объяснить*, что масса не зависит от скорости нелегко, а просто невозможно. Её можно так *определить*, а другой подход запретить, назвав архаичным и т.п., что студентов подчинит, думаю, однако, что временно. Демонстрация того, что механика Ньютона получается из эйнштейновской при $c \to \infty$, возможна и при использовании столь привычного огромному множеству физиков понятия массы, зависящей от скорости. Переход к $c \to \infty$ очевиден и в выражении для массы $m = m_0 / \sqrt{(1 - v^2 / c^2)}$ .

O2/A2/O1/A1$_{2.3}$. Нерелятивистское соотношение $p = mv$ просто неприменимо в релятивистском случае, где работает релятивистское соотношение $p = Ev / c^2$. Почему Вы считаете соотношение $p = mv$ "важным", "удобным и наглядным", если его в релятивистском случае просто нет? Почему нерелятивистский "хвост" должен "вилять" релятивисткой "собакой"?

*Why does the dog wag its tail?*
*Because the dog is smarter than its tail.*

*If the tail were smarter, it would wag the dog.*

Единственный ответ на мои вопросы, который я смог найти, заключается в том, что Вы являетесь профессионалом в области нерелятивистской физики, а лекции по теории относительности читаете, опираясь на свои студенческие воспоминания.

А3/О2/А2/О1/А1$_{2,3}$.. Могу лишь повторить, что соотношения $p = mv$ и $p = Ev/c^2$ при $E = mc^2$ просто тождественны. Тут нет ни «хвоста», ни «собаки». Мои студенты давно проявляли бы чудовищную отсталость, опирайся я лишь на «студенческие воспоминания». Просто я, помимо чтения разной литературы, твёрдо отрицаю принцип «всё старое и отжившее должно быть уничтожено в зародыше».

### Замечание А1$_3$ из письма Амусья А1

А1$_3$. Всегда отмечаю, что подстановка вместо массы покоя $m_0$ выражения $m = m_0 / \sqrt{(1 - v^2/c^2)}$ в любые формулы классической механики, без разбора, не делает их автоматически релятивистскими, как не делает подобная подстановка в уравнение Шредингера его релятивистским.

О1/А1$_3$. Очень хорошо, что Вы учите студентов тому, что нельзя автоматически подставлять соотношение между релятивистской массой и массой покоя в нерелятивистские уравнения. Но это должно использоваться Вами для того, чтобы объяснять им, что эти две массы являются историческими артефактами и замутняют понимание специальной теории относительности. Ведь масса, зависящая от скорости была введена ещё до создания теории относительности, именно для того чтобы сохранить нерелятивистскую связь импульса со скоростью. Почему Вы не хотите объяснять студентам, что релятивистская масса и масса покоя - это артефакты?

А2/О1/А1$_3$. Потому, что эти массы артефактами не считаю.

О2/А2/О1/А1$_3$. А почему Вы так не считаете? Ведь "релятивистская масса" и "масса покоя" - это действительно исторические артефакты! Или Вы не согласны с тем, что эти два понятия были введены в физику до создания теории относительности при безуспешных попытках согласовать между собой уравнения механики Ньютона и электродинамики Максвелла? Когда и кем они были, по-вашему, введены?

А3/О2/А2/О1/А1$_3$. СТО показало, что связь этих понятий, в определённом смысле эмпирическая, имеет глубокий физический смысл.

### Замечание А1$_4$ из письма Амусья А1

А1$_4$. Нередко отмечаю, что и эффективную массу, которая естественно возникает в теории многих тел, скажем, однородного электронного газа, нельзя подставлять вместо электронной массы в формулу для плазменной частоты. Такая подстановка грубо ошибочна.

О1/А1$_4$. Эффективная масса в случае электронного газа не имеет никакого отношения к обсуждаемому нами понятию массы свободной частицы. Зачем Вы упомянули её?

A2/O1/A1$_4$. Это убедительный пример того, что "автоматика" не работает в теоретической физике.

O2/A2/O1/A1$_4$. Я согласен с тем, что "автоматика" (т.е. бездумное использование формул) в теоретической физике не работет. Но как применить это к защищаемому Вами тезису о "релятивисткой массе" **свободной** релятивистской частицы? Причём тут плазма?

A3/O2/A2/O1/A1$_4$. Здесь это просто иллюстрация возможной ошибки, идущей от «автоматики».

### Замечание A1$_5$ из письма Амусья A1

A1$_5$. Широко пользуюсь техникой диаграмм Фейнмана в их нерелятивистской версии.

O1/A1$_5$. Фейнмановские диаграммы - замечательное орудие. В релятивистской физике они используют только понятие инвариантной массы и не используют понятие релятивистской массы и массы покоя. Зачем Вы ссылаетесь здесь на нерелятивистские диаграммы Фейнмана?

A2/O1/A1$_5$. Диаграммы Фейнмана и СТО не идентичны. Есть техника диаграмм релятивистская, есть - нерелятивистская.

O2/A2/O1/A1$_5$. Поскольку нерелятивистская физика являетя предельным случаем релятивистской физики, диаграммы Фейнмана работают и в нерелятивистской физике. Но обуждать с помощью нерелятивистских диаграмм Фейнмана вопрос об инвариантной массе релятивистской частицы невозможно. В релятивистские диаграммы Фейнмана входит лишь инвариантная масса, как для реальных, так и для виртуальных частиц. А в нерелятивистском пределе это не видно.

A3/O2/A1/5. Мого лишь повторить то, что хотел сказать – диаграммы Фейнмана и СТО не эквивалентны, хотя релятивистские диаграммы и базируются на СТО.

### Замечание A1$_6$ из письма Амусья A1

A1$_6$. Коллеги по еврейскому университету, которым я задал Ваш вопрос, преподающие СТО, возраста 40-60 и старше, пользуются одновременно двумя массами и двумя подходами, считая разницу между ними чисто терминологической.

O1/A1$_6$. Вы пишете, что Ваши коллеги по университету, как и Вы, пользуются двумя подходами к понятию массы. Согласны ли Вы переслать им это письмо и сказать, что я был бы им благодарен, если бы они объяснили мне, зачем и почему они это делают?

A2/O1/A1$_6$. Выше я уже упоминал про коллег. Их право действовать, как они считают нужным, используя методы и подходы, с их точки зрения наиболее доходчивые и удобные в том или ином курсе для студентов и т.д. Вашу позицию они видят чисто терминологической, каковой она является и по моему мнению.

О2/А2/О1/А1$_6$. Ни один из Ваших коллег не сообщил мне, зачем и почему пользутся понятиями "релятивистская масса" и " масса покоя". Они тоже специалисты в нерелятивистской физике?

А3/О2/ А2/О1/А1$_6$. Один из них – крупнейший специалист по ОТО. Несколько других – «полисты»[21]. Честно говоря, не вижу, зачем им Вам сообщать о причинах своих предпочтений. С Вашей и Л. Б. Окуня точкой зрения они знакомы не только через меня, но и из доклада Л. Б. Окуня на коллоквиуме в Еврейском университете Иерусалима, который он делали лет десять назад.

**Замечание А1$_7$ из письма Амусья А1**

**А1$_7$.** Ни я, ни они не видят "безобразия" формулы $E = mc^2 = m_0\gamma c^2$. Лично я работаю в атомной системе единиц и мне система единиц с $c = 1$ не удобна. Вообще, выбор специфической системы единиц - всё-таки дело профессионалов в данной области. Не думаю, что изложение релятивистской квантовой механики удобно начинать с выражения $\hat{p} = 1$ [22], которое, хотя и правильно, но очень трудно объяснять.

ОКУ: Для удобства ответа разобью Ваше замечание 1$_7$ на три – 1$_{7.1}$, 1$_{7.2}$ и 1$_{7.3}$

О1/А1$_{7.1}$. Вы пишете, что ни Вы, ни Ваши коллеги не видите безобразия формулы $E = mc^2$. Но ведь она не согласуется с основной формулой теории относительности для свободной релятивистской частицы $E^2 = m^2c^4 + p^2c^2$. Почему эта нестыковка не кажется вам безобразной?

А2/О1/А1$_{7.1}$. Вы не хуже меня понимаете, что здесь нет нестыковки. Просто Вы сознательно поставили в оба выражения одну и ту же массу, что неверно. Надо было писать $E^2 = m_0^2c^4 + p^2c^2$, всего и дел то.

О2/А2/О1/А1$_{7.1}$. "Всего и дел то" поставить с ног на голову преподавание специальной теории относительности, введя в него абсолютно не присущее ей понятие массы, зависящей от скорости, возникшее до создания этой теории. В общепринятой научной формулировке СТО масса есть релятивистский скаляр и потому не зависит от скорости. Вы серьёзно предлагаете переписать все статьи и книги по физике элементарных частиц, заменив в них $m$ на $m_0$, только потому, что в молодости Вы не выучили теорию относительности по книгам Ландау и Лифшица, или Ахиезера и Берестецкого, или Боголюбова и Ширкова, или Пескина и Шрёдера, или многих других авторов?

А3/О2/А2/О1/А1$_{7.1}$. Ничего в преподавании СТО не надо ставить с ног на голову. Напротив, не надо поднимать, как говаривал <u>А. Зиновьев</u>, «пустяк до уровня общегосударственной проблемы». Ничего не надо переписывать, как и не надо создавать иллюзию, будто Л. Б. Окунь, при полной поддержке коллег, поправляет отдельных заблудших. То, что им и Вами делается, будто борьба с ересью. Но ереси то и нет. А есть попытка простой терминологический вопрос поднять до уровня боьбы с опаснейшим заблуждением. Один, по-вашему, недоучил в молодости, другой – от старости недопонимает. Забавно, что *все*, кроме Пескина и Шрёдера, книги, перечисленные Вами, я читал, и ещё некоторые другие, но остался в многолюдном лагере «непонятливых». И спор наш меня не превращает в Вашего «прозревшего» адепта. А сколько лет Л. Б. Окунь, практически в полном одиночестве, воюет с этими «ошибочными» и «устаревшими» терминами, без, как показывает даже беглый просмотр ситуаци, заметных сдвигов?

---

[21] Специалисты по теории поля.

[22] Здесь $\hat{p} \equiv \sum_0^3 \gamma_\mu p^\mu$ есть скалярное произведение четырёхмерных векторов матриц Дирака $\gamma$ и энергии-импульса $p$.

А1$_{7.2}$ Лично я работаю в атомной системе единиц и мне система единиц с $c = 1$ не удобна. Вообще, выбор специфической системы единиц всё-таки дело профессионалов в данной области.

О1/А1$_{7.2}$. Лично Вам система единиц с $c = 1$ не удобна, так как Вы работаете в атомной системе единиц. Конечно, занимаясь атомной физикой, удобно пользоваться атомной системой единиц. Но занимаясь вопросом о соотношении массы и энергии, удобней пользоваться системой $c = 1$. А в этой системе равенство $E = mc^2$ выглядит как $E = m$. В результате у Вас возникает две разные буквы для энергии: $E$ и $m$, а для массы Вам приходится вводить $m_0$. Почему Вы не согласны с тем, что система $c = 1$ обнажает неуместность формулы $E = mc^2$ в преподавании теории относительности?

А3/О1/А1$_{7.2}$. Считаю, что само выражение «неуместность формулы $E = mc^2$ в преподавании теории относительности» неуместно. Хочу ещё раз подчеркнуть, что неверно полагать всю нерелятивисткую физику неким объектом меры нуль в общем множестве природных объектов. Чтобы убедиться в этом, достаточно взглянуть на распределение статей в таком всеостороние охватывающем всю физику журнале, как Physical Review (все серии). Да и не думаете ли Вы, что это больше Л. Б. Окуня и Ваша проблема, чем моя? Система с $c = 1$ неудобна не только мне, но и всем, работающим, в основном, в области нерелятивистской физики

О3/А3/О1/А1$_{7.2}$. Нет, я думаю, что проблема эта целиком Ваша. Вопрос о понятии массы в СТО - это по определению вопрос релятивистской физики. Поэтому обсуждать его в рамках нерелятивистской физики, как это всё время хотите Вы, невозможно. В релятивистской физике очень удобно непосредственно сравнивать энергию, импульс и массу, имеющих одинаковую размерность в единицах, где $c = 1$, но принадлежащих разным представлениям группы Лоренца. Энергия и импульс принадлежат векторному представлению, а масса - скалярному. А защищаемая Вами формула $E = mc^2$ приравнивает компоненту вектора скаляру.

А1$_{7.3}$. Не думаю, что изложение релятивистской квантовой механики удобно начинать с выражения $\hat{p} = 1$, которое, хотя и правильно, но очень трудно объяснять.

О1/А1$_{7.3}$. Согласен с Вами, что изложение релятивистской квантовой механики не следует начинать с уравнения Дирака $\hat{p} = m$ для частицы со спином 1/2. Его надо начинать с уравнения Клейна-Гордона-Фока $p^2 = m^2$ для частиц со спином нуль. Согласны ли Вы с тем, что эта формула в квантовой механике эквивалентна $E^2 - p^2 = m^2$?

А2/О1/А1$_{7.3}$. Всё-таки полноценным обобщением квантовой механики на движения со скоростями, близкими к световым, где сохранён смысл волновой функции, входит первая, а не вторая её производная по времени, является уравнение Дирака. Уравнение

КГФ писалось ведь не для π-мезона, которого тогда не было. В этом смысле, оно просто не верно. Но позднее и у него, с более глубокой трактовкой, нашлось приложение. Примеров того, что математика шире физического понимания и уравнение может оказаться справедливым там, где и не предполагалось - немало.

О2/А2/О1/А1$_{7.3}$. Вы серьёзно утверждаете, что уравнение Клейна-Гордона-Фока для бесспиновой частицы было просто неверно в течение 20 лет, пока не были открыты π-мезоны? Чем уравнение КГФ неполноценно? Обратите внимание, что уравнение $E^2 - p^2 = m^2$ появилось сначала в квантовой механике и только через 15 лет (в 1941) в классической теории поля.

А3/О2/А2/О1/А1$_{7.3}$. Более точно: уравнение Клейна-Гордона-Фока было беспредметно и безобъектно, и именно поэтому, в отличие от уравнения Дирака, не стало предметом награждения Нобелевской премией.

**Замечание А1$_8$ из письма Амусья А1**

ОКУ: Для удобства ответа разобью Ваше замечание 1$_8$ на три – 1$_{8.1}$, 1$_{8.2}$ и 1$_{8.3}$

А1$_{8.1}$. Система единиц $\hbar = c = 1$ кажется мне не очень удобной для обычных студентов, не аспирантов, поскольку затрудняют переход к классике и нерелятивизму. Не могу согласиться с Вашим утверждением, что "$c$ в качестве единицы скорости навязывается самой природой, когда мы хотим рассматривать фундаментальные процессы в ней". Всё-таки это дело соглашения.

О1/А1$_{8.1}$. Почему Вы не согласны с тем, что скорость света навязывается нам природой при рассмотрении фундаментальных вопросов? Какую другую скорость Вы предлагаете использовать вместо $c$? В каком смысле скорость света - "это дело соглашения"?

А2/О1/А1$_{8.1а}$. Почему "избранность" скорости света заставляет её полагать единицей - понять не могу. Как и избранность постоянной Планка. Это вопрос удобства в области занятий, но также и механизм понимания теории.
А2/О1/А1$_{8.1б}$. Удобно видеть явно и $\hbar$ и $c$, если хочу иметь возможность перехода к классике и нерелятивизму.
А2/О1/А1$_{8.1в}$. Кстати, чисто гипотетически, что Вы будете делать, если мир окажется не СРТ инвариантным[23], и найдутся взаимодействия, не Лоренц-инвариантные, пусть и крайне малозаметные? Что ещё предложите положить равным единице? А ведь до открытия электродинамики физике вполне хватало галилеевой инвариантности. Неужели развитию физики наступил конец, и навсегда появилась объективно «избранная привилегированная система единиц»?

О2/А2/О1/А1$_{8.1а}$. Вы не предложили никакую другую скорость вместо $c$. Но вся СТО основана именно на "избранности" $c$. А вся квантовая механика основана на "избранности" $\hbar$. Квантовая теория поля и релятивистские диаграммы Фейнмана основаны на "избранности" $\hbar$ и $c$. Электрический заряд характеризует силу взаимодействия в электродинамике, константа Ньютона - в гравитации. Есть ещё и другие константы, характеризующие сильное и слабые взаимодействия. Но все они в этом смысле менее "избранны", чем $\hbar$ и $c$.

---

[23] имеется в виду инвариантность законов физики относительно изменения знака зарядов (C), пространственной (P) и временной (T) чётности.

А3/О2/А2/О1/А1$_{8.1a}$. Избранность скорости света не *заставляет* (как пишете Вы), а *позволяет* полагать её равной единице при интерпретации СТО. Это удобно при рассмотрении очень быстрых движений и в частности обсуждаемого нами вопроса о том, зависит ли масса от скорости СТО и квантовая механика не есть замкнутые науки, но части физики. Поэтому, на мой взгляд, при их изложении уместно руководствоваться интересами всей этой науки, а не только того или иного, пусть и очень важного, её раздела. Здесь вполне уместно использованное Вами выражение про то, что хвосту не следует вилять собакой.

О2/А1$_{8.1в}$.Ваше утверждение, что скорость света - "это дело соглашения" напомнило мне о статье А. А. Тяпкина, которую он в 1960х годах пробивал через ЦК КПСС в УФН, а в 1971 пробил. На её публикации настаивал В. Л. Гинзбург, против её публикации выступали Я. Б. Зельдович и Л. Б. Окунь. Я знаю, что в 1971 году он ездил на квартиру к В. А. Фоку в высотном здании на площади Восстания, где Фок при нём написал резко отрицательную рецензию на эту статью (имею копию этой рецензии, которую прилагаю вот она).
«В Редакцию Журнала "Успехи Физических Наук"
В Отделение Ядерной Физики
В Отделение Общей Физики и Астрономии

О Т З Ы В   о статье А.А.ТЯПКИНА
ВЫРАЖЕНИЕ ОБЩИХ СВОЙСТВ ФИЗИЧЕСКИХ ПРОЦЕССОВ В ПРОСТРАНСТВЕННО-ВРЕМЕННОЙ МЕТРИКЕ СПЕЦИАЛЬНОЙ ТЕОРИЙ ОТНОСИТЕЛЬНОСТИ

По-видимому, автор не сомневается в правильности теории относительности как физической теории. Но в то же время он считает неудовлетворительной её общепризнанную формулировку на основе преобразований Лоренца и предлагает вернуться к преобразованиям Галилея, вместо принятых в теории относительности преобразований времени, имеющих прямой физический смысл и составляющих вместе с преобразованием Лоренца для координат известную группу Лоренца, автор предлагает пересчитывать время во всех системах отсчёта на какое-то условное единое время (соответствующее одной какой-то системе отсчёта).

Предлагаемая автором трактовка формально возможна (и даже тривиальна, поскольку обще-ковариантный аппарат допускает произвольные преобразования), но чрезвычайно громоздка и только затемняет физический смысл всех соотношений.

Автор возвращается к старым высказываниям создателей теории относительности и, произвольно толкуя их, пытается доказать, что они не понимали или неправильно понимали основы теории. При этом автор допускает неточности в формулировках (например, смешивает преобразования Галилея и принцип относительности Галилея).

Автору чужда мысль, что правильная формулировка законов природы должна быть простой и изящной, и он не отмечает и не рассматривает фундаментальной роли симметрии (группы преобразований в пространстве Минковского), важной не только при описании электромагнитных явлений.

Резюмируя, можно сказать, что статья А. А. Тяпкина не содержит каких-либо новых идей, заслуживающих внимания. То, что он говорит и предлагает - многословно громоздко и не нужно, а то, что там верно, то тривиально. Читателя, не знакомого с теорией относительности (если таковые есть через 65 лет после её создания), статья может только запутать. Читателю же, знакомому с теорией, статья ничего положительного не даёт.

Печатать подобную статью в Успехах Физических Наук ни в коем случае нельзя. Напечатание её серьёзно подорвало бы репутацию журнала.

Академик

30 сентября 1971 г.

/В.Фок/»

Тяпкин утверждал, что в СТО надо пользоваться не преобразованиями Лоренца, а преобразованиями Галилея. А то, что при этом скорость света в прямом и обратном направлении различна, не останавливало его. Он так же как и Вы, считал константу $c$ "делом соглашения". Статью Тяпкина см. УФН **106** (4) 618-659 (1972). Там же на стр. 660-662. Статья Кадомцева, Келдыша, Кобзарева, Сагдеева. Какую из этих двух статей Вы считаете правильной?

A3/O2/A1$_{8.1в}$. Это удивительное совпадение, но, полемизируя с Вами, я тоже вспомнил Тяпкина. Он старательно объяснял, как «неправильно» понимали классики свои собственные работы. Именно, делая открытия, они не отваживались указать на «своё место» выскочке – Эйнштейну. Всё это Тяпкину было нужно, чтобы найти иных авторов теории относительности. И А. А. Логунов утверждал, явно намеренно игнорируя суть вопроса, что создателем СТО был не Эйнштейн, а Пуанкаре. Но дело в физике не сводится к формулам и просто их преобразованиям. Введение понятия массы, зависящей от скорости, т. е. записанное в любой, но выбранной системе координат, понятно иллюстрирует достижение СТО. Предшествующие, доэйнштейновские, попытки её введения в этом смысле к делу не относятся.

Кстати, присланная Вами рецензия Фока не спорит с *возможностью* пользоваться «единым временем» Тяпкина, но отмечает, что предлагаемое тем - многословно и не нужно, а то, что у него верно, то тривиально. Я увидел в «едином времени» Тяпкина нечто подобное единой массе – просто не удобно потребителю - пользователю.

Разумеется, в споре Кадомцева. Келдыша, Кобзарева и Сагдеева с Тяпкиным я на их стороне.

O2/A1$_{8.16}$. Разумеется, при переходе к нерелятивистскому пределу удобно выписать скорость света $c$ явно. Но к обсуждаемому нами вопросу о зависимости массы от скорости в релятивистском пределе, это отношения не имеет.

O2/A1$_{8.1в}$. В 20 веке произошли две великие революции - релятивистская и квантовая. Система единиц $\hbar c = 1$ позволяет лучше понять смысл каждой из них, понять смысл квантовой механики и теории относительности. И в частности, понять, почему неразумно говорить о том, что масса зависит от скорости.

На основе $\hbar c$ построена вся наша цивилизация, столь непохожая на цивилизацию времён Галилея. Ожидаете ли Вы чего-либо даже отдалённо подобного от теоретических спекуляций о нарушении СРТ и Лоренцовой инвариантности?

Согласны ли Вы с тем, что наша с Вами задача учить людей жить в этом релятивистском и квантовом мире, а не вешать им на уши лапшу столетней давности?

А3/О2/А2/8.1в. Не согласен с такими утверждениями. Не могу, да и не хочу гадать, что дадут те или иные спекуляции в теоретической физике, да и более земные подходы, вроде теории струн. Будет суждено, увидим. Про «лапшу столетней давности» – законы Ньютона старше, но верх неразмия вводить их студентам предельным переходом $c \to \infty$ от СТО или, тем более, ОТО.

О1/А1$_{8.2}$. Согласны ли Вы с тем, что систему $\hbar c = 1$ надо преподавать наряду с CGS, и что тогда никаких трудностей при переходе к классике и нерелятивизму у студентов не будет?

А2/О1/А1$_{8.2}$. Совершенно с Вами не согласен. Считаю просто неверным. Этот выбор системы единиц крайне затруднит переход к классике и нерелятивизму.

О2/А2/О1/А1$_{8.2}$. Поясните, пожалуйста, почему Вы думаете, что знание системы $\hbar c = 1$ затруднит, например, переход к нерелятивизму. В уравнении $E^2 - p^2 = m^2$ очень легко восстановить степени $c$, если вспомнить, что в механике Ньютона размерность импульса равна произведению размерностей массы и скорости, а размерность энергии равна произведению размерностей массы и квадрата скорости. И сразу получаем $E^2/c^4 - p^2/c^2 = m^2/c^2$. Почему Вы думаете, что студенту не следует этого знать? Знание системы единиц и выбор системы единиц - это две разные вещи. Знание системы $\hbar c = 1$ в нашем релятивистском и квантовом мире необходимо студенту-физику. А выбор её в нерелятивистском пределе свидетельствовал бы, что он не выдержал экзамена.

Отвергая фундаментальный характер константы $c$ как скорости света, Вы демонстрируете, что расхождение между нами не только в терминологии, но и в том, что у нас совершенно разные картины мира.

А3/О2/А2/О1/А1$_{8.2}$. Приписывая мне отрицание фундаментальной роли скорости света в современной картине мира, Вы просто отклоняетесь от истины, да и превратно толкуете написанное мною. Но, выучившись на прошлом, физик обязан допускать, что дальнейшее изучение природы внесёт нами не предвидимые поправки в сегодняшнюю ясность. Если бы этого не произошло – тут крылся бы абсолютно новый, фундаментальнейший из законов природы.

## Замечание А1$_9$ из письма Амусья А1

А1$_9$. Упоминаемое как непоследовательность или объясняемое Вами иными причинами сосуществование у Фейнмана в его книгах, Ландау (в книгах с Лифшицем или Румером), да и у самого Эйнштейна обоих подходов к массе, считаю проявлением их отношения к вопросу как к чисто терминологическому.

О2/А1$_9$. Ландау и Фейнман были величайшими педагогами и выбору адекватной терминологии уделяли большое внимание. У Ландау и Лифшица есть только инвариантная масса, в научных трудах Фейнмана есть только инвариантная масса. Как мне удалось выяснить, книжечка Ландау и Румера была написана ещё в тридцатых годах, до 37 года. Она, возможно, была написана под сильным влиянием Эйнштейна, с которым Румер работал на стыке 20-х и 30-х годов. Об истории этой книги Л. Б. Окунь пишет в "`Релятивистской кружке" (см. ссылку выше). Мне писал один из авторов "Фейнмановских лекций по физике" - Сэндс, что Фейнман знал эту книжечку. Возможно, что она повлияла на него, когда он продумывал свои лекции.

А2/О2/А1$_9$. Эти доводы мне напомнили то, чем пользовался профессор Васильев в давно мною слушанной лекции по парапсихологии. Согласных с собой он повышал в научных званиях, делая их "академиками", а несогласных - понижал. Или другой пример. Человек проводит расчёт по своей идее, причём часть результатов совпадает с данными опыта, а часть - нет. И автор скрупулёзно объясняет причину несогласия, считая согласие самоочевидным подтверждением своей теории.

Даже само название "книжечка" по отношении к написанному Ландау и Румером, приведенная Л. Б. Окунем в "Релятивистской кружке" шутка в адрес этой работы, ссылка на 1937 г., должны создать у читателя предвзятое отношение к этой работе. Во влиянии работавшего у Эйнштейна Румера на Ландау, по принципиальному вопросу – сомневаюсь. Влияние в принципе популярной книги Ландау на суждения Фейнмана по серьёзному, с точки зрения последнего, вопросу, считаю совершенно невероятным, имея в виду научный стиль и известный характер самого Фейнмана.

О3/А2/О2/А1$_9$. Я не понял, почему я Вам напомнил парапсихолога. Скорей, его методика должна напоминать Вам Вашу. Ведь это Вы, пренебрежительно назвав проблему "чисто терминологической", пытаетесь уравнять в правах массу, зависящую от скорости, и инвариантную массу.

Ласкательное слово "книжечка" просто отражает тот факт, что в этой маленькой книжке карманого формата было около 70 страниц. Вы, по-видимому, спутали его с пренебрежительным словом "книжонка"?

Румер и Ландау написали эту книжечку до ареста их обоих. И так они думали о массе до 1937 года. В 1941 году Ландау и Лифшиц нашли общепринятый теперь способ введения инвариантной массы. Но Ландау и Румер для широкого круга читателей в конце 50х годов сохранили старый текст с релятивистской массой, который оии не стали менять. Ясно, что в это время публикация была важной для Румера, а Ландау не отнёсся к ней серьёзно. Что касается Фейнмана, то я не вижу, почему надо не доверять свидетельству Сэндса. Почему Вы ему не доверяете?

А3/О3/А2/О2/А1$_9$. Дело не в недоверии к кому-нибудь, а в сомнении правдоподобия Вашей гипотезы о влиянии книги Ландау и Румера в вопросе, который Фейнман считал бы принципиальным и важным.

## Замечание А1$_{10}$ из письма Амусья А1

А1$_{10}$. Фраза Ландау об его с Румером книге как о месте, где "два жулика уговаривают третьего, что за гривенник он может понять, что такое теория относительности" отнюдь не отражает, на мой взгляд, отношения Ландау к общей книге. Ведь третьего - читателя, можно полагать невеждой, но не жуликом. Скорее, речь идёт, если такое высказывание и имело место, об общих трудностях популяризации науки.

О1/А1$_{10}$. У меня нет оснований не верить Румеру, когда он пишет о трёх жуликах. Конечно, популяризаторы обманывают читателя, создавая у него иллюзию понимания. Но если Ландау 70 лет тому назад понял, что масса, зависящая от скорости - это жульничество, то какие сегодня могут быть основания продолжать жульничество?

А2/О1/А1$_{10}$. Звучит неубедительно. Популяризаторов - два, а говорится о трёх жуликах. Как доверчивый читатель может считаться жуликом, или Ландау и здесь, в счёте до трёх, допустил ошибку? Кроме того, из шутки Ландау в пересказе Румера совершенно не следует, что "Ландау 70 лет тому назад понял, что масса, зависящая от скорости - это жульничество". Здесь очевидная натяжка с Вашей стороны.

О2/А2/О1/А1[10]. Ландау не ошибся, назвав жуликом "доверчивого читателя", котрый за гривенник хочет понять, что такое СТО. За гривенник СТО не понять. 70 лет тому назад Ландау и Лифшиц поняли, что надо объяснять СТО тем, кто действительно хочет понять эту теорию. Это экспериментальный факт, а не натяжка, и пересказа Румера здесь нет. Я просто сопоставил этот факт с шуткой Ландау.

А3/О2/А2/О1/А1[10] Не вижу жульничества в поведении читателя, тем более, что во времена, когда Ландау говорил о «трёх жуликах», плату за обучение взимать было не принято. Что-то иное он имел в виду. Да это и не важно. Моя цель лишь отметить, что построив в 1941 г. изложение СТО без явного введения понятия массы, зависящей от скорости, он и в конце 50-х не видел в её использовании криминала. Он мог иронизировать над собой и другим, но сбывать «неверное или отжившее» не стал бы во имя помощи Румеру. Да и конец 50-х – не такое время, когда Румера надо было спасать, выпуская неверную книжку. Словом, сам Ландау мешает причислению к библейским своего с Лифшицем текста 1941 г. Оно и понятно: возможность не есть необходимость или обязательность.

## Замечание А1[11] из письма Амусья А1

А1[11]. Вот и Л. Б. Окунь, говоря об энергии атома, делит энергию входящего в него электрона на кинетическую и потенциальную, и вводит скорость атомного электрона, определённым значением которой он, как известно, не обладает.

О1/А1[11]. Вы упрекаете Л. Б. Окуня в том, что он говорит про скорость электрона в атоме. Думаю, что он имел в виду модуль скорости. В каком месте, какой его статьи надо это оговорить?

А2/О1/А1/11. В статье 1989 г. в УФН

О2/А2/О1/А1[11]. Перечитав эту и другие свои статьи на эту тему, я вижу, что Вы правы. При стандартном изложении квантовой механики (через волновую функцию) эти понятия скорости, кинетической и потенциальной энергии электрона в атоме заменяются соответствующими операторами. Надо было обяснить, что в данном квантовом состоянии, у электрона в атоме есть только энергия связи, а определённых значений скорости, потенциальной и кинетической энергии у него нет. В своей новой книге "Азы физики" Л. Б. Окунь это подробно разъясняет.

А3/О2/А2/О1/А1[11]. Прежде операторов, я бы упомянул соотношения неопределённостей и дуализм «волна-частица». Хотя понятия и старые, они, вследствие противоречия каждодневному опыту, вновь и вновь волнуют поколения людей. Примечательно, что и сегодня квантовая механика, в особенности её копенгагенская интерпретация, под такой же атакой, как и, например, в 50-е. И скрытые параметры стараются ввести, и подходы а-ля Бом. Говорят, будто копенгагенская трактовка устарела, используют для доказательства квантовой механики неравенства Белла, как-будто нет мириад фактов в её поддержку без неравенств Белла.

Сам в июле 2010, в Турине, на конференции «открытый форум европейской науки» (Euro Science Open Forum) участвовал в дискуссии на эту тему, которую организовал и вёл Г. 'тХоофт (G. 't Hooft). Напоминало диспут 50-х в СССР, с участием философов и просто непричастных. Я выступал с консервативных, копенгагенских позиций чуть ли не в одиночестве.

Думаю, что помимо или перед операторным подходом стоит описывать физику, стоящую за дуализмом, хотя это и может казаться «лапшой столетней давности» по сравнению с операторным подходом. Но ведь речь идёт об «Азах физики», а математизация создаёт иллюзию возможности и другого аппарата для описания той же физической реальности.

## Замечание А1₁₂ из письма Амусья А1

A1$_{12}$. Не очень нагляден и образ античастицы, как частицы, движущейся вспять во времени, предложенный, но не развитый, кстати, впервые Я. И. Френкелем.[24]

О1/А1$_{12}$. Согласен с Вами, что образ частицы, движущейся вспять во времени, не очень нагляден. Но сознаёте ли Вы, что на этом понятии основывается весь механизм релятивистских фейнмановских диаграмм? Это ключевое понятие в современной теории элементарных частиц. Ведь без этого нельзя рисовать фейнмановские петли.

А2/О1/А1$_{12}$. Конечно, это ключевое понятие, как и дырка в нерелятивистской теории многих тел.

О2/А2/О1/А1$_{12}$. Но дырка в релятивистской теории бозонов не работает, а петля работает. Рисовать фейнмановские петли без движения вспять во времени невозможно. Сознаёте ли Вы это?

А3/О2/А2/О1/А1$_{12}$. Дырку, как и петлю, вполне законно рассматривать и в бозонной теории. Роль «вспять во времени», разумеется, сознаю.

## Замечание А1₁₃ из письма Амусья А1

A1$_{13}$. Мне кажется, что многократное повторение в двух последних статьях одного подхода как "единственно правильного" не будет способствовать его распространению.

О1/А1$_{13}$. Но я так настаиваю на точной терминологии в отношении массы, потому что считаю её единственно правильной. Я понимаю, что моя бескомпромиссность должна вызывать раздражение у тех, кто привык к безответственной неточности. Как, по-вашему, должно довести свою точку зрения до читателей?

А2/О1/А1$_{13}$. Думаю, что это дело безнадёжное. Во-первых, Л. Б. Окунь и Вы свою точку зрения считаете единственно правильной. Но это не довод для других. Само обвинение других в "безответственной неточности" содержит субъективную, на непреложных научных фактах не основанную, оценку.

О2/А2/О1/А1$_{13}$. Точка зрения, которую я считаю единственно правильной, не является моей личной точкой зрения. Её придерживается большинство физиков, работающих в физике элементарных частиц. Только они, как правило, думают, что объяснить её "нерелятивистам" нельзя, а я убеждён, что можно, и свидетельством этого является наша переписка.

---

[24] О1/А$_{12}$. Буду благодарен, если Вы пришлёте ссылку на Френкеля. А2/О1/А1$_{12}$. Сейчас не могу этого сделать. По-моему, это во френкелевой Электродинамике (или Квантовой механике?) О2/А2/О1/А1$_{12}$. Пришлите, пожалуйста, эту ссылку. Этих книг у меня нет.

А3/О2/А2/О1/А1[13]. Но работает в физике элементарных частиц абсолютное меньшинство физиков. Да и они, в чём убедился в ходе прямых с ними разговоров, далеко не все – ваши и Л. Б. Окуня адепты.

**Замечание А1[14] из письма Амуся А1**

А1[14]. Не подходят, на мой взгляд, для дискуссии в УФН слова Л. Б. Окуня "сорняк массы, зависящей от скорости", или сопоставление использованного Э. Неизвестным в памятнике Хрущёву белого и чёрного мраморов с отношением к двум видам массы. Не годится и однозначное отнесение "массы покоя" к словам-паразитам. Если же говорить о лингвистике, то любой язык содержит "слова-паразиты", без которых языковые пуритане хотят обойтись. У них в итоге получается нежизнеспособная канцелярщина.

О1/А1[14]. Думаете ли Вы, что "Теория поля" Ландау и Лифшица, не использующая слов-паразитов "масса покоя" и "релятивистская масса" является образцом "нежизнеспособной канцелярщины", и что в неё надо добавить эти слова?

А2/О1/А1[14]. Совсем не обязательно. А вот название словосочетаний "масса покоя" и "релятивистская масса" "словами-паразитами" содержит необоснованную предвзятую, личную оценку.

О2/А2/О1/А1[14]. Но в перечисленных мною выше монографиях авторы прекрасно обходятся без этих слов-паразитов и адекватно излагают релятивистскую физику.

А3/О2/А2/О1/А1[14]. Не устану повторять, что нет и речи о «словах-паразитах», но лишь об удобстве последовательного изложения, избираемого авторами в каждом случае сообразно излагаемому материалу и адресуемой читательской аудитории.

**Замечание А1[15] из письма Амуся А1**

А1[15]. Не вижу, что в 20 веке шла "борьба двух концепций массы". Преувеличение это сильное.

О1/А1[15]. Какое же это преувеличение, если в 2011 году Вы посылаете мне Ваше письмо с 25-ю пунктами в защиту релятивистской массы?

А2/О1/А1[15]. Я писал "Не вижу, что в 20 веке шла "борьба двух концепций массы". Преувеличение это сильное". В моём ответе нет элемента борьбы концепций, поскольку этих двух концепций нет - есть вопрос об удобстве терминологии.

О2/А1[15]. Более ста лет продолжается борьба сторонников двух концепций массы. Одни пытаются сохранить за массой дорелятивистскую роль меры инерции и протащить в преподавание теории относительности нерелятивистскую формулу для импульса. Другие настаивают на том, что масса - это релятивистский инвариант и потому от скорости не зависит. Третьи пытаются сидеть на двух стульях сразу.

А3/О2/А1₁₅. Как уже писал выше, нет в реальности никакой «столетней войны» двух концепций массы. Нет в реальности никаких «двух стульев», а есть, при последовательном преподавании, вопрос удобства. Увы, есть и эмоциональные обороты про то, как «одни пытаются ... протащить», а другие, рыцари точности выражений, с ними борются. Вообще, слова типа «протащить» вместо «использовать» терпимы в политических дебатах, но задевают в том, что должно быть обсуждением чисто научно-методического вопроса.

**Замечание А1₁₆ из письма Амусьи А1**

А1₁₆. Задевающе звучат Ваши слова "*Пришла пора прекратить обманывать всё новые и новые поколения школьников и студентов* (курсив мой - МА), внушая им, что возрастание массы с увеличением скорости - это экспериментальный факт". Думаю, что нормальные преподаватели и научные работники - не невежды, и никого не обманывают. Также задевает Ваша фраза: "*Мне представляется, что недопустимо выдавать зависимость массы от скорости за экспериментальный факт, скрывая от студента* (курсив мой - МА), что она является интерпретационным фактоидом".

О1/А1₁₆. А Вы думаете, что зависимость массы от скорости - это экспериментальный факт, а не фактоид?

А2/О1/А1₁₆. При сохранении определения массы соотношением $p = mv$ - это факт. Понятие энергии тоже требует определения. Оно, кстати, вполне заметно эволюционирует, и неизвестно, наступил ли уже в этой эволюции последний шаг, в свете, простите за антикаламбур, чёрной материи и квинтэссенции.

О2/А2/О1/А1₁₆. Напомню, что фактоид похож на факт, но считается достоверным только потому, что встречается в печатных текстах. В теории относительности нерелятистской формулы для импульса тоже является фактоидом и встречается лишь в устаревших текстах. Именно потому, что физика эволюционирует, неразумно в начале 21 века сохранять в ней заблуждения конца 19 века и старательно прививать их всё новым поколениям.

Что же касается Вашего антикаламбура, то первая не астрономическая, а физическая статья про тёмную (зеркальную) материю была опубликована И. Я. Померанчуком, И. Ю. Кобзаревым и Л. Б. Окунем в рамках стандартной СТО в 1966 г. И последующие многочисленные авторы, обсуждавшие зеркальную материю, нарушения СТО тоже не предполагали. А космологический член (одной из разновидностей которого является квинтэссенция) появился при создании ОТО на полвека раньше. Так что ни СТО, ни ОТО эволюции физики не препятствовали.

А3/О2/А2/О1/А1₁₆. Не о том речь, что они мешали, а о том, к примеру, что некоторая модификация уравнений ОТО, проведенная проф. Я. Бекенштейном из Еврейского университета в Иерусалиме, позволяет обойтись без введения чёрной материи, да и, насколько я понимаю, квинтэссенции[25]. Следовательно, на сегодняшний день существование этих материй ещё модельно – зависимо.

---

[25] Тёмная материи и тёмная энергии (квинтэссенция) – компоненты Вселенной, составляющие свыше 20% и 70%, соответственно, всей её материи. На обычную материю, построенную из кварков, глюонов и лептонов приходится всего чуть более 4%. Экспериментальных фактов, позволяющих понять свойства этих объектов вселенной, крайне мало. С другой стороны, полагаясь на описание Вселенной в рамках

**Замечание А1$_{17}$ из письма Амусья А1**

А1$_{17}$. Отмечу, что если сказать школьнику или студенту, начинающему изучать физику, будто согласно СТО энергия нарастает с увеличением скорости, то это вызовет у него недоумение: "Причём тут СТО?" Ведь рост энергии с увеличением скорости следует и из известной с этак 7го класса школы формулы $E_{kin} = m\upsilon^2 / 2$.

О1/А1$_{17}$. Почему рост энергии тел с ростом их скорости должен вызывать недоумение у студентов и школьников? Им надо разъяснить, что при приближении скорости тела к скорости света этот рост становится гораздо более быстрым, чем в 7мом классе.

А2/О1/А1$_{17}$. Не считаю это правильным, ведь им сразу надо ввести и полную энергию покоя $E = m_0 c^2$, что довольно трудно, и выглядит голословно.

О2/А2/О1/А1$_{17}$. Вы настойчиво продолжаете использовать $m_0$ для обозначения массы и не хотите использовать $E_0$ для обозначения энергии покоя. (Фок писал $W_0 = mc^2$.) Но если не объяснить школьникам, что теория относительности содержит принципиально важное (новое 100 лет тому назад) понятие "энергия покоя", то они никогда не поймут эту теорию. При скорости частицы, приближающейся к скорости света, её энергия линейно растёт с ростом импульса и как $(1 - \upsilon^2 / c^2)^{-1/2}$ с ростом скорости.

А3/О2/А2/ О1/А1$_{17}$. Считаю, что школьников ничуть не меньше обогатит и понимание массы покоя, и обоснованное введение массы, зависящей от скорости. Разумеется, в любом варианте люди знакомятся с новым для них релятивистским корнем $(1 - \upsilon^2 / c^2)^{-1/2}$. Кстати, в применение к квантовой механике, великолепно сосуществуют две её формы – матричная и использующая въявь дифференциальные уравнения, поначалу казавшиеся существено разными.

**Замечание А1$_{18}$ из письма Амусья А1**

А1$_{18}$. Вы отмечаете, что Дж. А. Уилер много сделал для установления "понятия лоренц-инвариантной массы". Соответствующей ссылки у Вас не нашёл.

О1/А1$_{18}$. Вы просите ссылку на Уилера. Вот она: E. F. Taylor, J. A. Wheeler, Space-time Physics, Second edition, 1992, pp 246-252. Dialogue: Use and abuse of the concept of mass. В конце этого диалога Тейлор и Уилер пишут, что их позиция совпадает с моей.

А2/О1/А1$_{18}$. Посмотрю при первой возможности.

О2/А2/О1/А1$_{18}$. Посмотрели? Пришлю Вам этот диалог. Вставьте его, пожалуйста, в нашу дискуссию в этом месте.

---

уравнений ОТО Эйнштейна, без введения тёмная материи и тёмной энергии (квинтэссенции) не обойтись.

А3/О2/А2/О1/А1[18]. Ещё не посмотрел. А упомянутого диалога не нашёл. Его включение в наш спор было бы уместно. (См. «Диалог» в Приложении 2, а комментарий Амусья к нему - ниже).

**Замечание А1[19] из письма Амусья А1**

А1[19]. Считаю запись $m^2 = e^2 - p^2$ неудобной, несмотря на упоминание Л. Б. Окунем того факта, что рекламный щит на Рублёвке содержит формулу Эйнштейна с малым $e$. Предпочитаю сохранить $e$ как обозначение заряда электрона, а $E$-энергии.

О2/А1[19]. Я тоже предпочитаю $E$, но в статье Л. Б. Окуня в Am J Phys объясняется, что иногда удобней $e$. Забавно, что Вы не готовы использовать букву $e$ для энергии и заряда, но используете букву $m$ для энергии и массы.

А2/О2/А1[19]. Я не использую букву $m$ для энергии и массы. Считаю систему с $c = 1$ для преподавания не специализирующимся на релятивистской теории поля и физике высоких энергий - не подходящей.

О3/А2/О2/А1[19]. Выше мы уже обменялись мнениями о системе с $c = 1$. Повторюсь. Если Вы преподаёте только нерелятивистскую механику, то система $c = 1$ не нужна. Если объясняете СТО, она необходима.

А3/О3/А2/О2/А1[19]. И в интересующих меня релятивистских задачах, и при преподавании пользуюсь по привычке атомной системой единиц. Все пока целы: и студенты, и я.

**Замечание А1[20] из письма Амусья А1**

А1[20]. С. У. Хокинг не просто популяризатор теории относительности. Он заслуживает упоминания в контексте обсуждаемого вопроса не только из-за того, что говорится на суперобложке русского издания его книги. Той, что переведена на 40 языков. Для Л. Б. Окуня существенно, что книга пропагандирует формулу $E = mc^2$. Но ведь это не криминал, а свидетельство определённой точки зрения, хотя от Вашей с Л. Б. Окунем и отличающейся.

О1/А1[20]. Это свидетельство того, что вопрос им не продуман.

А2/О1/А1[20]. Не согласен. Могу лишь повторить написанное ранее.

О2/А2/О1/А1[20]. Это криминал, поскольку на 40 языках он пропагандирует формулу, которой нет в теории относительности.

А3/О2/А2/О1/А1[20]. Не устану повторять, что Ваше утверждение – «нет в теории относительности» – суть несомненное преувеличение, по отношению к специалисту масштаба Хокинга, определённо имеющего право считать, что есть такое теория относительности и вчера, и сегодня.

**Замечание А1[21] из письма Амусья А1**

A1$_{21}$. Не возьму в толк, в противоречии с тем, что Вы пишите, как представление о релятивистской массе может помешать пониманию механизма передачи всей энергии быстро движущегося электрона покоящемуся. Иное дело, как ведёт себя с ростом энергии вероятность такого процесса.

О1/А1$_{21}$. А Вы можете себе представить, как энергию тяжёлой биты передать мячику от пинг-понга? Причём здесь вероятность?

А2/О1/А1$_{21}$. Задача о столкновении быстрого электрона с медленным решается, как Вы отлично знаете, элементарно, и в амплитуде явно прослеживается вклад процесса с передачей большой энергии и почти без передачи - за счёт неразличимости электронов. Бита и мячик от пинг-понга к делу отношения не имеют.

О2/А2/О1/А1$_{21}$. Известный физик, задавший этот вопрос, просто хотел применить (защищаемую Вами) точку зрения, что масса релятивистского электрона намного больше, чем масса покоящегося (бита и шарик). Неразличимость электронов прямо доказывает, что эта точка зрения неправильна. Согласны?

О2/А2/О1/А1$_{21}$. Не надо, как уже говорилось, даже известному физику использовать тот факт, что масса зависит от скорости, без всяких ограничений. Как сказал один старый профессор-медик своей студентке, правда, по иному, поводу: «Нельзя, милочка, мои слова воспринимать столь буквально». Ведь так ненароком можно опровергнуть и принцип неразличимости электронов, поскольку, имея разные скорости и массы, они, тем самым, вполне различимы.

## Замечание А1$_{22}$ из письма Амусья А1

А1$_{22}$. На мой взгляд, это очень хорошо, что в учебники и научные статьи не ввели обсуждаемые Л. Б. Окунем понятия "инерции" или "инера", "тяжести" или "грава". От этого путаницы было бы невповорот. Побольше, чем от "конгруентных" треугольников и квадратов там, где в них нет ни нужды, ни квалифицированных кадров для разъяснения.

О1/А1$_{22.1}$. Согласен, что все эти термины вводить не надо, надо просто не приписывать массе понятия меры инерции и меры гравитации, как это делал Ньютон и временами Эйнштейн.

А2/О1/А1$_{22.1}$. Приходится объяснять смысл значков в формулах. Например, что входит в $F = ma$ и в $F \sim mM / r^2$. Даже записав релятивистские выражения для $E$ и $p$, приходится их связывать с привычными из обыденной жизни и нерелятивистской физике энергией и импульсом. Без такой связи уравнения, например, сохранения лишаются содержательности - что сохраняется, собственно? Просто компоненты некоторого 4-х мерного вектора?

О2/А2/О1/А1$_{22.1}$. Конечно, надо объяснять, как измеряются энергия и импульс. И надо объяснять, как из релятивистсих (точных) формул для энергии и импульса следуют нерелятивистские (приближённые) формулы для них. И не надо объявлять формулу $p = mv$ точной.

О1/А1$_{22.2}$. Какое высказывание Л. Б. Окуня о конгруентных треугольниках Вам не понравилось?

А2/О1/А1$_{22.2}$. Не Л. Б. Окуня или Ваше, а колмогоровское[26], при его революции преподавания математики школьникам.

О2/А2/О1/А1$_{22.2}$. Мне кажется, что Бурбаки не годятся школьникам, так как излишне усложняют элементарную математику. То, как Л. Б. Окунь излагает СТО, упрощает формулы, а не усложняет их.

А3/О2/А2/О1/А1$_{22.2}$. Упрощение формул, несомненно важно. Помню, как нам уравнения гидродинамики записывали в компонентах, или в «Магнитном электроне» де Бройль приводил в компонентах, а не в компактной форме, уравнения Дирака. Но о том, что видывал эти уравнения в сложной записи - не жалею: это позволило прочувствовать их сложность. Упрощение формул, важное само по себе, особенно существенно, когда проясняет физический смысл. Пример – рассматривая СТО, можно распологать системы отсчёта так, что одна из осей сонаправлена относительной скорости. Соответствующие формулы для преобразований Лоренца резко упрощаются.

## Замечание А1$_{23}$ из письма Амусья А1

А1$_{23}$. Не замечал ничего в поддержку утверждения Л. Б. Окуня "Забавно, что обычно те, кто сравнивает механику Ньютона с астрологией, верят, что масса зависит от скорости". Обратное определённо неверно - вижу это у себя и многих коллег.

О1/А1$_{23}$. Это сделал А. Г. Сергеев, переводивший книгу Хокинга "Мир в ореховой скорлупе". В марте 2008 года он опубликовал в журнале "`Вокруг света" статью "`Почему мы доверяем науке?". Можете ли Вы прочесть эту статью?

А2/О1/А1$_{23}$. Автора не знаю, если он сделал такое утверждение - значит, явно допустил ошибку. Зачем мне его читать? Напомню, о чём шла речь. Я писал: «Не замечал ничего в поддержку утверждения Л. Б. Окуня "Забавно, что обычно те, кто сравнивает механику Ньютона с астрологией, верят, что масса зависит от скорости". Обратное определённо неверно - вижу это у себя и многих коллег».

О2/А2/О1/А1$_{23}$. Обратного утверждения Л. Б. Окунь не делал, а прямое Вы не опровергаете.

А3/О2/А2/О1/А1$_{23}$. Явно нелепое высказывание переводчика едва ли уместно использовать для подкрепления своеей точки зрения. Оно, скорее, способно её скомпрометировать.

## Замечание А1$_{24}$ из письма Амусья А1

А1$_{24}$. Думаю, что в изложении многих областей физики, включая и такую важную, как СТО, уместно руководствоваться известным стишком

---

[26] Имеется в виду школьные учебники математики, в создании которых важнейшую роль сыграл А. Н. Колмогоров.

*Мой милый, тебе эту песню дарю.*
*Рассчитывай силы свои*
*И если сказать ты не можешь "хрю-хрю",*
*Визжи, не стесняясь, "и-и!".*

Относя себя к "визгливым поросятам", я ии-каю, оказываясь, при этом отнюдь не в самой плохой компании - старых, и не только, специалистов в обсуждаемой области. Конечно, мода и воззрения меняются,

*и то, что было ясность мудреца,*
*Потомкам станет бредом сумасшедших…*

Но замечательно написал Дюрренматт: "Нет ничего сомнительнее, чем вера, которая подавляет сомнение".

O1/A1$_{24}$. Мне кажется, что преподавая СТО, не надо ни хрюкать, ни визжать. Надо просто пользоваться адекватной терминологией. А высказывание Дюренмата, на мой взгляд правильное, но относится оно не ко мне, а к Вам.

A2/O1/A1$_{24}$. Так я и считаю широко используемую методу, в том числе, с зависимостью массы от скорости - вполне адекватной, если это делать грамотно, а не суя, куда ни попадя, релятивистскую массу. Я сослался: "Но замечательно написал Дюрренматт: "Нет ничего сомнительнее, чем вера, которая подавляет сомнение". Всё-таки это в ваш и Л. Б. Окуня адрес, ибо я не насаждаю "единственно правильную" точку зрения, вопреки мнению явно довольно многих специалистов-физиков касательно единственности этой правильной точки зрения.

O2/A2/O1/A1$_{24}$. Вы называете верой профессиональное знание СТО, а я называю верой пропаганду неадекватной терминологии.

A3/O2/A2/O1/A1$_{24}$. Профессиональное знание явно было у Фока, есть у Хокинга, да и у многих других физиков, не разделяющих максималистскую или абсолютистскую, уже этим несколько не гармонирующую с теорией относительности, точку зрения Л. Б. Окуня. Разумеется, неадекватной терминологию людей, с Л. Б. Окунем и Вами не согласных, не считаю.

## Замечание A1$_{25}$ из письма Амусья A1

A1$_{25}$. Полемизируя с Л. Б. Окунем, я, наверное, провоцирую резкую отповедь, типа "ты, Зин, на грубость нарываешься". Но свои точки зрения скрывать не привык. Да и возраст виляния не поощряет. Кроме того, ещё раз обращаю Ваше внимание на то, с чего начал - с расползания анти- и псевдо-науки и того, как там используются научные работы и критика признанно великих, о чём говорил и что имел ввиду в своём исходном письме.

O1/A1$_{25}$. Вы пишете о Вашем возрасте. Я не смог найти в сети Вашей биографии. Сколько Вам лет? Кто Ваши учителя и ученики?

А2/О1/А1$_{25}$. Мне семьдесят седьмой год. Найти мою биографию в интернете нет проблем - и на сайте ФТИ[27], и на сайте Hebrew University Jerusalem Amusia. Кончал, если не удастся Вам прочесть про это в сети, кафедру физики Ленинградского Университета. ОТО слушал и сдавал Фоку, СТО - Друкарёву, математику - Смирнову, и т.д. См., например http://www.phys.huji.ac.il/~amusia/. На сайте же найдёте имена учеников, занимающихся сейчас широким спектром проблем – проблемой трёх тел, астро- и биофизикой, атомами, фуллеренами, эндоэдралами, кластерами, и т.д. Всего 25 кандидатов, из них 12 – доктора наук. Там же – места их работы.

О2/А2/О1/А1$_{25}$. Большое спасибо. Прочёл Вашу биографию. Очень благодарен Вам за возможность обсудить с Вами спорные вопросы. Вы обвиняете меня и Л. Б. Окуня в том, что наши статьи способствуют расползанию анти-науки. Я не могу с Вами согласиться. Мы стараемся излагать предмет как можно точней. Я уверен, что если бы Вы перестали употреблять привычные Вам двойные стандарты в отношении массы и тоже стали пользоваться адекватной терминологией, то это привело бы к уменьшению анти- и псевдо- науки вокруг нас.

А3/О2/А2/О1/А1$_{25}$. Я писал: "Обращаю Ваше внимание на то, с чего начал -- с расползания анти- и псевдо-науки и того, как там используются научные работы и критика признанно великих, о чём говорил и что имел ввиду в своём исходном письме". Здесь, как видно, нет и речи о том, что используемая Л. Б. Окунем и Вами ковариантная запись СТО антинаучна. Речь идёт о том, что написанное Л. Б. Окунем, сдобренное в адрес излагающих по-другому и самих этих изложений набором крайне резких замечаний, создаёт иллюзию бурной дискуссии, оставляя впечатление, что сами физики запутались в "трёх соснах". Значит, думает "вечный двигатель" или ниспровергатель Максвелла, «и мне всё можно». Не имея ни знаний, ни идей. Я таких встречал, и немало.

О3/А3/О2/А2/О1/А1$_{25}$. Я внимательно прочёл все работы Эйнштейна по теории относительности (прочли ли Вы их?) и очень высоко ценю его вклад в науку. Но что делать, если он не всегда бы прав? Л. Б. Окуня и мои статьи направлены не против Эйнштейна, а против тех, кто пытается канонизировать его ошибки и оговорки, как это делал, например, Джеммер.

А4/О3/А3/О2/А2/О1/А1$_{25}$. Критиковать и устранять ошибки можно и нужно не только у Эйнштейна, но и у Ландау с Померанчуком. Однако мы с Вами, на мой взгляд, никаких ошибок Эйнштейна не обсуждали, а говорили про то, как лучше излагать его идеи и результаты научной и околонаучной публике.
      Я кое-что читал Эйнштейна, иногда по необходимости обращался и к немецким оригиналам. Прочтение привело, например, к выводу, возможно и неправильному, поскольку досконально, на уровне рукописей и личных писем вопрос не исследовал, что называть частицы бозонами крайне несправедливо. Действительно, Бозе в посланной им Эйнштейну работе говорит лишь об электромагнитном излучении, а вот Эйнштейн сразу не исключает и возможности других объектов новой статистики.
      Прочёл я внимательно и работу Эйнштейна 1905 г. об излучении, обычно именуемую статьёй про фотоэффект. Там я увидел явно незамеченное процитированным Вами секретарём Шведской академии содержание. На тему этой работы в 2005 г. сделал ряд докладов и собираюсь, кстати, написать статью в УФН. Но

всё не доходят руки. Возможно, дискуссия с Вами подхлестнёт. На мой взгляд, эта статья явно недооценена и её место во всём цикле работ Эйнштейна, выключая создание СТО, должным образом не понято.

О4/А4/О3/А3/О2/А2/О1/А1$_{25}$. Выше мы обсуждали уравнение Эйнштейна $E_0 = mc^2$, которое он иногда записывал в виде $E = mc^2$ и которое многие интерпретируют как рост массы с ростом скорости. Эта ошибочная интерпретация затрудняет изучение теории относительности.

А5/О4/А4/О3/А3/О2/А2/О1/А1$_{25}$. Не считаю, что указанная интерпретация ошибочна и противоречит СТО. Никаких доказательств *ошибочности* не прочёл.

Жду подробных ответов на все мои вопросы. Их последняя версия приведена синим.
Желаю Вам всего доброго. Ваш Леон Болеславович

Отвечаю зелёным. Старался, как мог.
С наилучшими пожеланиями, всегда Ваш, Мирон Я. Амусья.

05.11.11

Глубокоуважаемый Леон Болеславович
Ещё раз просмотрел «Переписку Леона Болеславовича ОКУ и М. Я. Амусья о зависимости массы от скорости в специальной теории относительности (СТО)», начатую письмом М. Я. Амусья 31.03.11 (А1) и законченную 16 июня 2011 и могу лишь повторить написанное 18 сентября с. г.: «Мне кажется, что уже много усилий затрачено с Вашей стороны на изложение и обоснование Л. Б. Окуня и Вашей точки зрения, а с моей стороны - на спор с Вами. Позиции разъяснены. Думаю, что в целом получился интересный материал, заслуживающий обнародования в Архивах или на Трибуне УФН. Дальнейшая полемика приведёт лишь к повторам». Я спросил вас тогда, что Вы думаете об этом. Ответа на свой вопрос не получил, хотя рекомендованное Вами ознакомление с книгой Л. Б. Окуня и страницами из Тейлора-Уилера подтвердили, в моих глазах по крайней мере, уместность приведенного выше предложения.
Это не означает, что я намереваюсь переписку с Вами прервать, но Тейлор и Уилер – не прямые участники нашей полемики, а книга Л. Б. Окуня – гораздо шире вопроса о том, вводить или нет массу покоя. Поэтому то, что я пишу, есть, на мой взгляд, начало новой переписки.
Разумеется, избранные страницы Тейлора и Уилера ближе к теме[28]. С них и начну. Оба автора – известные физики, особенно Уилер, секретарь Н. Бора и учитель Р. Фейнмана. Оба известнейшие и опытнейшие преподаватели. Словом, мне опять приходиться «замахиваться» вверх. И тем не менее... Хотя бы потому, что Уилер знал по крайней мере одну мою работу и на неё ссылался, соглашаясь, я позволю себе сослаться, не соглашаясь. Наверное, более правильно было бы делать это в прямой переписке с авторами, но, по меньшей мере, с одним из них это, увы, невозможно.
Приведенное Тейлором и Уиллером изложение в целом мне не понравилось, начиная с термина momenergy, который взывает к введению ещё и dadenergy. Эти слова имели бы замечательный перевод – мамуля-энергия и папуля-энергия – стр. 246. Не понял я, по прочтении рекомендованных Вами стр. 246-252 (см. Приложение), в чём состоит abuse (злоупотребление) концепции массы. Показалось мне, что ими самими вокруг mass (масса) создан mess (ералаш, беспорядок).

---

[28] См. Приложение

Не знаю, кому адресована книга, но, как писал ранее, система единиц с $c = 1$ для общего учебника не очень удобна, а для подготовленного читателя большинство вопросов на стр. 246-252 просты и имеют очевидные ответы.

Авторы на стр. 247 убеждают читателя, что понятие momenergy богаче, чем просто массы. Это очевидно, поскольку и в «старом» подходе масса покоя, равно как и масса, зависящая от скорости, не содержит информации о направлении движения частицы. Словом, очевидно, что четыре числа содержательнее одного или даже двух.

Странно звучит совет на стр. 248 о том, как найти массу системы слипшихся частиц. «Взвесьте её!», - советуют авторы. А как же ещё? Обнюхать или нагреть, что ли? Но, с другой стороны, и сами понятия held – or – stick together (держать – или слипаться), нуждаются в уточнении. К примеру, не всякому студенту ясен ответ на вопрос, будет ли муха, назойливо вьющаяся около банки мёда, не садясь на него, влиять на результат взвешивания. Хотя она тоже в каком-то смысле stick.

«Таким образом, часть массы *превратилась* в энергию, но *масса* системы не изменилась», - пишут авторы на стр. 249, что трудно отнести к легко понимаемым утверждениям.

После этого странно звучит, будто эйнштейновское утверждения об эквивалентности массы и энергии не означает, что это одно и то же. Авторы всюду теперь используют слово «масса» вместо «масса покоя». Ясно, что возникают непонятности с эквивалентностью, утверждаемой Эйнштейном. Но все эти непонятности целиком «сделаны» авторами.

Знаменитое, а теперь, усилиями «модернизаторов» ставшее злополучным, соотношение $E = mc^2$ работает не только в системе покоя, если считать, что $m = m_0 / \sqrt{(1 - v^2 / c^2)}$.

Концепция «релятивистской массы» порождает, как утверждают авторы на стр. 250, недопонимание. Именно, «она делает увеличение энергии объекта со скоростью или моментом, представляющимся связанным с какими-то изменениями во внутренней структуре объекта». По-моему, это просто неверно. Несомненно, и это прямо следует из самого принципа Лоренц - инвариантности, что изменение энергии при переходе от одной инерциальной системы к другой есть кинематический эффект, а не отражение динамических процессов в системе.

На стр. 251 авторы объясняют, почему понятие «массы покоя» ими сейчас отбрасывается. Как они каются, это понятие было в первом издании их книги. Теперь, под влиянием «вдумчивого студента», они поняли, что с «массой покоя» что-то не так. «Что происходит с «массой покоя» частицы, когда она движется?», - повторяют они вслед за студентом. Каков вопрос, такой и ответ: «В действительности, масса есть масса есть масса. Она имеет одинаковое значение во всех системах координат, она инвариантна вне зависимости от того, как частица движется». Очень понятно, и, главное, убедительно, не правда ли? К делу привлечён и покойный, задолго до создания теории относительности, великий Галилей, сказавший: «В вопросах науки авторитет тысячи не стоит скромного размышления одного», что, конечно, правильно, но к делу прямо не относится. Очевидно, что и тысяча далеко не всегда неправа.

Совсем непонятна картинка на стр. 251, где, наглядно интерпретируя соотношение $E^2 = m_0^2 + p^2$ с помощью прямоугольного треугольника, авторы изображают гипотенузой не $E$ а $m$!

Констатирую: вендетта, направленная против «массы покоя» и релятивистской массы, опять оказалась голословной, а потому – безрезультатной.

Свои комментарии о книге Окуня отложу, если позволите, до следующего письма.

С наилучшими пожеланиями, всегда Ваш, Мирон Я. Амусья.

29.11.11

Глубокоуважаемый Леон Болеславович!

Начну с повтора. Думаю, что нашему с Вами спору можно и нужно придать гласность. Уместно, думаю, начать с эпиграфа, в качестве чего предлагаю следующий короткий диалог:

> *-Как вам удалось переубедить столь многих*
> *маститых предшественников?*
> *-А я их не переубеждал. Я их просто пережил.*
> Беседа студента со знаменитостью

Внимательно прочитал книгу Л. Б. Окуня «АЗЫ ФИЗИКИ. Очень краткий путеводитель». Теперь могу высказать свои впечатления. В целом, мне она понравилась и показалась полезной. Встречающаяся в нескольких местах критика понятия массы, как величины, в рамках СТО зависящей от скорости, мало отражается на содержании. Скажу так – она его не портит.

Затрудняюсь отнести книгу к определённому жанру – это не учебник, не монография, это, скорее, шпаргалка для «так сказать, профессоров»[29], которые вознамерились прочитать курс всей физики студентам, и полезно напомнить, что стоит туда включить. Как и предполагал, книга убедительно показала, что парой хлебов можно накормить множество людей. Однако, тогда же отмечал, идя по стопам старого анекдота, что они едва ли будут сыты. Так и оказалось. Но, тем не менее, налицо кормёжка. Приятно, что смена изложения не привела к использованию системы единиц с $c = \hbar = 1$.

Печально, что в малый объём книги не удалосьне уложить «неизбежность странного мира», т.е. пояснение того, что изменения в его картине, вносимое физикой, не есть свободная игра воображения, но результат вынужденного ухода от простейших решений и схем – под напором новых теорий и опытов или наоборот. Именно знание причин этой неизбежности мешает «вечным двигателям» совершать перевороты в науке.

Думаю, что упоминание «бегущих констант»[30], наряду с истинными константами, мешает пониманию, поскольку в действительности масса и заряд начинают отличаться от своих пустотных значений при учёте диаграмм более высоких порядков. Замечу, что в теории многих тел эти изменения имеют вполне понятный физический смысл, приводя, например, к понятию квазичастицы и эффективной массы $m_{eff}$ в теории ферми-жидкости Ландау. Под перенормируемостью теории понимаю взаимную компенсацию высших расходящихся (бесконечных) поправок к линиям частиц и вершинам взаимодействий. Это позволяет их суммировать и положить $m_{eff} = m_{exp}$ и $e_{eff} = e_{exp}$.

---

[29] Выражение из первой версии книги Л. Б. Окуня, которую Вы мне прислали.

[30] Имеется в виду изменение электронных массы и заряда с уменьшением расстояния между взаимодействующими частицами за счёт поляризации ваккуума.

К сожалению, не нашёл в книге Л. Б. Окуня идеи асимптотической свободы[31], хотя она вполне попадает в нужные и яркие азы.

Отмечу, что, строя картину мира на константах $c$ и $\hbar$, Л. Б. Окунь, начиная с раздела 16.1, говорит о её ограниченности из-за существования того, что можно было бы назвать Планковским миром, с его массой, длиной, временем, определяемыми ещё и гравитационной постоянной $G_N$, которая, выходит, тоже вполне фундаментальна для понимания мира, и в её физическом смысле стоит разобраться ничуть не менее, чем в $c$ и $\hbar$ (см., однако, стр. IV «Азов физики»).

Это правильно, что Л. Б. Окунь отнёс Главы 24-26 к спекулятивным. Хотелось бы, однако, услышать мнение, пусть и спекулятивное, столь крупного эксперта, по другим, кроме высоких энергий, разделам физики. Однако, перестав заботиться об обосновании «неизбежности странного мира», когда открывается кажущееся абсолютно вольным фантазирование, стоит помнить, что патологи – «вечные двигатели» начинают чувствовать себя вольготно. Раз «всё дозволено», чем они хуже остепенённых и имеющих высокие звания профессионалов? Открывать им дорогу плохо, а потому всегда надо отвечать на вопрос, даже в спекулятивных главах, перефразирующий классика: «Чего хочу, с какою целью, я, люди, это вам пишу? Какому злобному веселью кретинам повод подаю?».

Теперь перейду к замечаниям, не имеющим отношения к нашему спору (I), затем вернусь к нашему спору (II). В конце упомяну неточности и опечатки (III).

## I. Замечания, не имеющие отношения к нашему спору

1. На мой взгляд, вся физика не равна физике элементарных частиц. В книге же Л. Б. Окуня, этой части физики, по ходу изложения, уделяется всё большее внимание, полностью вытесняя в конце остальное – физику твёрдого тела, конденсированного состояния, сверсильных полей. Конечно, есть ограничения по объёму, но кое-что про элементарные частицы спокойно можно было бы опустить.

2. Несбалансированно приведены имена авторов открытий – некоторые есть, часто – не упомянуты даже Нобелевские лауреаты. А имя автора открытия, с помощью Википедии, есть ресурс, и значительнейший, для дополнительного чтения.

3. Есть достойные, но не упоминаемые Л. Б. Окунем идеи. Например, идея возможной зависимости констант $\hbar$, $e$ и $c$ от времени, которая обсуждается специалистами, начиная с Дирака. Это тем более важно, что, как пишет Л. Б. Окунь на стр. iv, «не усвоив их физического смысла, в потоке информации и дезинформации, обрушиваемых на нас интернетом, разобраться нельзя». Признаюсь, однако, что смысла самой фразы я не понял.

4. Думаю, это неправильно считать (см. 4.5, стр. 11 книги Л. Б. Окуня), что «основной вклад в создание СТО внесли последовательно Лоренц, Пуанкаре, Эйнштейн и Минковский». СТО есть физическая теория **одного** автора. Даже первые двое в списке Л. Б. Окуня были весьма далеки от понимания смысла СТО как необходимой теории преобразования реальных пространственных координат и времени. Очень важная геометризация СТО, проведённая Минковским, вообще относится к аппарату и удобству представления, но не к физической сути теории.

5. Общее замечание по линиям на диаграммах Фейнмана. Фотонные линии, раз ось время идёт слева направо, не стоит ориентировать строго вертикально. Это

---

[31] Имеется в виду ослабление взаимодействия между кварками с уменьшением расстояния между ними. Интуитивно, на основе опыта в макро-, да и микромире, мы привыкли к ситуации, когда взаимодействие между частицами ослабевает с ростом расстояния между ними, а с уменьшением расстояния – растёт. рос

выглядит как мгновенность передачи взаимодействия. Это же относится и к фермионным линиям.

6. В диаграммной теории многих тел так называемому морю Дирака соответствует набор заполненных в основном состоянии уровней, вплоть до фермиевского. Удобно ввести и вакуумное море бозонов – например, в теории сверхпроводимости или сверхтекучести. С этой точки зрения мнение Фейнмана о «море Дирака» мне понятно.

**II. Замечания по предмету нашего спора о зависимости массы от скорости.**

1. На стр. ii Л. Б. Окунь пишет, что «уравнение $E = mc^2$, интерпретируемое как зависимость массы частицы $m$ от скорости $v$, даёт ложное представление о том, что такое масса. Закрепляется это ложное представление тем, что массу называют массой покоя и обозначают $m_0$, что разрывает связь между теорией относительности и механикой Ньютона, являющейся предельным случаем теории относительности при малых скоростях». Это неверное по существу и никак не следующее из самой книги Л. Б. Окуня совершенно голословное утверждение.

2. Приводя в 4.7 уравнения $\vec{v} = \vec{p}c^2 / E$ (4.9), Л. Б. Окунь пишет: «С ростом полной энергии тела прирост его скорости становится всё меньше, и она стремится к предельной величине $c$». Прямо увидеть это из (4.9) гораздо сложнее, чем из осуждаемой им, но правильной формулы $E = mc^2 = m_0 c^2 / \sqrt{1 - v^2 / c^2}$. Из неё непосредственно следует, что для конечности $E$ при $v \to c$ должно быть $m_0 \to 0$. Л. Б. Окунь пишет: «Замечательным свойством формул (4.8) и (4.9) является то, что они применимы и к массивным, и сколь угодно лёгким частицам, и к частицам с массой, равной нулю». Очевидно, что эти слова равно применимы и к $E = m_0 c^2 / \sqrt{1 - v^2 / c^2}$: для конечности $E$ при $v \to c$ должно быть $m_0 \to 0$.

3. В Разделе 27.1 «Солнце всходит и заходит, а в тюрьме моей темно». Там, в этой тюрьме зловредный профессор просто массово портит студентов и студенток, он «не проповедует им поверхностную правду» про БАК[32], где масса протона возрастает в тысячи раз, а «закладывает ложные понятия, основанные на пресловутой формуле $E = mc^2$, в представления своих студентов об основах теории относительности». Неприемлемо грубая эта фраза бездоказательна и неправильна по существу, о чём я многократно сообщал Вам в нашей переписке.

4. В рамках Раздела 27.2 не понял, в чём проявляется сознание всё большего числа людей того, что наша цивилизация квантовая и релятивистская, а не информационная, например.

5. Раздел 27.5 не имеет отношения ни к сути, ни к истине в науке, а содержит лишь сетования о том, что не все «делают, как я». Так может быть дело во мне, т.е. в Л. Б. Окуне? Напоминая знаменитое «Карфаген должен быть разрушен», звучит аккордное сетование о «неспособности многих профессиональных физиков признать, что в специальной теории относительности масса не зависит от скорости». По счастью, физика – не римский сенат, да и оснований у Катона было много больше. Что касается якобы всеобщего восприятия квантовой механики как науки, делающей мир призрачно зыбким, то с этим согласиться тоже не могу. Любой студент-физик знает, что не будь законов квантовой механики, электроны бы попадали сразу на свои ядра, ликвидировав нас с Л. Б. Окунем да и землю, луну и т.д.

---

[32] БАК – Большой адронный коллайдер, построенный на границе Швейцарии и Франции.

**III. Неточности и опечатки**

1. Слову «учёный» (стр. iV) предпочитаю «научный работник». А кот – он учёный.

2. Преобразования Галилея есть преобразования координат и времени в классической механике при переходе от одной инерциальной системы к другой. Уравнение Л. Б. Окуня (4.2) $\vec{r}' = \vec{r} + \vec{v}t$ связывает два положения тела с интервалом времени $t$ в одной и той же инерциальной системе. Поэтому (4.2) есть просто уравнение равномерного движения.

3. *В 4.1 фраза «Все частицы данного типа абсолютно тождественны*[33] *и имеют одну и ту же массу m» неудачна, поскольку не определяет слово «тип», и не отмечает различие квантовой неразличимости от классической.*

4. В 5.2 «Симметричный тензор, построенный из двух векторов, *как нетрудно видеть*, имеет пять компонент 3х3=1+3+5» - понять, не зная, невозможно

5. В 5.4 Л. Б. Окунь возвращается к *тождественности*, и наряду с «типом» появляется и «сорт».

6. Перед 8.1 фраза «потенциальная энергия электрона не зависит от его кинетической энергии» странна, поскольку связанный электрон в поле протона имеет определённой только полную энергию.

7. В 9.5 конфигурация атомов иная, нежели указано в 9.4: $^{36}Kr(3d^{10}, 4s^2, 4p^6)$ и $^{54}Xe(4d^{10}, 5s^2, 5p^6)$.

8. В противовес сказанному в Разделе 10.1 образование молекул объясняется не обязательно обменом электронами.

9. В Разделе 10.4 стоило отметить, что, согласно квантовой механике, $E \neq 0$ при $T = 0$.

10. В Разделе 11.4 стоит указать, что реальные частицы приходят (уходят) не только с (на) бесконечно далёких расстояний, но и времён.

11. В 16.2, если фотону приписать массу $m = \hbar\omega / c^2$, действие на него поля гравитации понимается весьма просто. Кстати, подобное качественное объяснение облегчит понимание Раздела 16.1.

12. В Разделе 18.3 стоит уточнить, что в $E / mc^2$ раз возрастает время жизни мезона в лабораторной системе координат, а не просто время жизни.

13. Говоря в Разделе 19.8 по сути о «восьмеричном пути»[34], Л. Б. Окунь не упоминает израильского теоретика Ю. Неемана, на этом пути опередившего М. Гелл-Манна, т.е. опубликовавшего работу ранее его. Отмечу, что и в предложении частиц с дробными зарядами (1/3 и 2/3 заряда электрона), позднее названных кварками, Нееман с сотрудником был впереди. Ими «кварки», однако, трактовались как чисто математические элементы. Впрочем, считал их реальными частицами первоначально лишь Дж. Цвейг.

14. В конце Разделов 20.1 и 20.2 зачем-то повторяется «симметрия нарушается в слабых процессах на сто процентов». Несправедливо, говоря о несохранении чётности, опускать имя Ч.-С. Ву в Разделе 20.2.

15. Раздел 22.1 создаёт ощущение произвола, а не «неизбежности странного мира», что выглядит не очень хорошо.

16. В конце Раздела 22.6 уместно было бы написать чётко о роли приближённых симметрий и причинах их нарушений.

---

[33] Курсив всюду мой – М. А.
[34] «Восьмеричный путь» - своего рода таблица Менделеева для элементарных частиц.

17. В Разделе 22.7 утверждается: «из-за того, что протон легче нейтрона следует, что *u*-кварк легче *d*-кварка». Это не убеждает, поскольку у них могут быть несколько разные взаимодействия, чтобы объяснить разницу в 1.3 Мэв между протоном и нейтроном.

18. В Разделе 23.10 уместно было бы упомянуть: речь идёт о встречных пучках, что позволяет резко выиграть в энергии столкновения.

19. В Главе 24 стоило бы упомянуть далеко не измученных известностью и популярностью авторов идеи суперсимметрии.

20. В Разделе 26.1 говорится, что с начала 1980-х годов *большинство теоретиков*, занимающихся физикой на планковской шкале, работают над теорией суперструн». «Большинство не всегда право»,- как говаривал В. И. Ленин, да и наука – не очередь за колбасой, так что где большинство – абсолютно неважно.

С наилучшими пожеланиями, всегда Ваш, Мирон Я. Амусья.

28.01.12

Глубокоуважаемый Леон Болеславович!

В соответствии с Вашей рекомендацией, прочитал исправленный вариант «Азов физики», и комментирую проведенную Л. Б. Окунем «работу над ошибками». Так я называю уточнения и поправки, которые внесены в ответ на мои замечания или, вполне возможно, вне всякой связи с ними. Думаю, «теперь твоя душенька довольна?», как говаривал некий старик своей знаменитой старухе. Напомню, что дискуссия, на мой взгляд, получилась, и заслуживает быть представлена «городу и миру». Полного чуда не произошло, и каждая из сторон в основном осталась на своих позициях. Но конкретных замечаний у меня сейчас гораздо меньше.

Конечно, сохранилось несогласие во всём, что касается «бичевания» Л. Б. Окунем зависимости массы от скорости. Разумеется, повторение даже бездоказательного утверждения придаёт ему дополнительную силу. Как в знаменитом телефонном разговоре из старого анекдота: «Рабинович, вы дурак! Кто говорит? Все говорят!». Но научная работа, по счастью, не сводится просто к учению, которому повторение – мать. Она требует и доказательств…

Не повторяю замечаний, учёт которых требовал дополнительного места и времени – Л. Б. Окунь ответил на него в Постскриптуме 2. Но, тем не менее, остались конкретные замечания, которые можно было учесть, не тратя места и особого времени.

**I. Замечания, не имеющие отношения к нашему спору**

7. Общее замечание по линиям на диаграммах. Фотонные линии, раз ось время идёт слева направо, не стоит ориентировать строго вертикально, что выглядит как мгновенность передачи взаимодействия. Это же относится и к фермионным линиям.

**II. Неточности и опечатки**

1. Преобразования Галилея есть преобразования координат и времени в классической механике при переходе от одной инерциальной системы к другой. Ваше уравнение (4.2) $\vec{r}' = \vec{r} + \vec{v}t$ связывает два положения тела с интервалом времени $t$ в одной и той же инерциальной системе. Поэтому (4.2) есть просто уравнение движения.

2. В 9.5 конфигурация атомов иная, нежели указано в Разделе 9.4: $^{36}\text{Kr}(3d^{10}, 4s^2, 4p^6)$ и $^{54}\text{Xe}(4d^{10}, 5s^2, 5p^6)$ - она определяется потенциалом ионизации подоболочки.

3. Говоря в Разделе 19.8 по сути о «восьмеричном пути», вы не упоминаете израильского теоретика Ю. Неемана, на этом пути опередившего М. Гелл-Манна, т.е. опубликовавшего работу ранее его. Отмечу, что и в предложении частиц с дробными зарядами (1/3 и 2/3 заряда электрона), Нееман с сотрудником был впереди. Ими «кварки», однако, трактовались как чисто математические элементы. Впрочем, считал их реальными частицами первоначально лишь Цвейг.
4. В конце 20.2 несправедливо опускать имя Ву.
5. В Главе 24 стоило бы упомянуть далеко не измученных известностью и популярностью авторов идеи суперсимметрии.

С наилучшими пожеланиями Вам и Л. Б. Окуню, всегда Ваш, Мирон Я. Амусья.

**Постскриптум**

Думаю, что это письмо завершит нашу переписку. Она стала привычной, и мне будет не хватать Ваших требований и возражений. Но единожды начатое, должно иметь и конец ранее, чем обратится в простой повтор и дурную бесконечность.

**Пост-постскриптум 03.02.12**

Только что вновь прочитал в Фейнмановских лекциях по физике, т.1, Введение: «Могут ли все-таки результаты опыта быть ошибочными? Могут, из-за нехватки точности. Например, масса предмета кажется неизменной; вращающийся волчок весит столько же, сколько лежащий на месте. Вот вам и готов «закон»: масса постоянна и от скорости не зависит. Но этот «закон», как выясняется, неверен. Оказалось, что масса с увеличением скорости растет, но только для заметного роста нужны скорости, близкие к световой. Правильный закон таков: если скорость предмета меньше 100 км/сек, масса с точностью до одной миллионной постоянна. Вот примерно в такой приближенной форме этот закон верен. Можно подумать, что практически нет существенной разницы между старым законом и новым. И да, и нет. Для обычных скоростей можно забыть об оговорках и в хорошем приближении считать законом утверждение, что масса постоянна. Но на больших скоростях мы начнем ошибаться, и тем больше, чем скорость выше». Молодец Ричард Фейнман. Отлично подытожил наш с Вами спор, Леон Болеславович!

**Приложение**

**E. F. Taylor, J. A. Wheeler**, *Space-time Physics*, Second edition, 1992, pp 246-252. Dialogue: "Use and abuse of the concept of mass".

# DIALOG: USE AND ABUSE OF THE CONCEPT OF MASS

Does an isolated system have the same mass as observed in every inertial (free-float) reference frame?

Yes. Given in terms of energy E and momentum $p$. $m^2 = E^2 - p^2$ in one frame, by $m^2 = (E')^2 - (p')^2$ in. another frame. Mass of an isolated system is thus *invariant*.

Does its *energy* have the same value in every inertial

No. Energy is given by $E - (m^2 + p^2)^{1/2}$ or frame?

$$E = m/(1 - v^2)^{1/2}$$

or

$$E = (mass) + (kinetic\ energy) = m + K$$

Value depends on the frame of reference from which the particle (or isolated system of particles) is lib-served. Value is lowest in the frame of reference in which the particle (or system) has zero momentum (zero *total* momentum in the case of an isolated system of particles). In that frame, and in that frame only, energy equals mass.

Does energy equal zero for an object of zero mass, such as a photon or neutrino or graviton?

No. Energy has value $E = (0^2 + p^2)^{1/2} = p$ (or in conventional units $E_{conv} = cp_{conv}$). Alternatively one can say—formally—that the entire energy resides in the form of *kinetic* energy ($K = p$ in this special case of *zero* mass), none at all in the form of rest energy. Thus,

$E = (mass) + (kinetic\ energy) = 0 + K = K = p$

(case of zero mass only!)

Can a photon — that has no mass –give mass to an absorber?

Yes. Light with energy E transfers mass $m - E$ (= $E_{conv}/c^2$) to a heavy absorber (Exercise 8.5).

Invariance of mass: Is that feature of nature the same as the principle that all electrons in the universe have the same mass?

No. It is true that all elementary particles of the same kind have the same mass. However, that is a fact totally distinct from the principle that the mass of an isolated system has identical value in whatever free-float frame it is figured (invariance of system mass).

Invariance of mass: Is that the same idea as the conservation of the momenergy of an isolated system?

No. Conservation of momenergy — the principle valid for an isolated system — says that the momenergy 4-vector figured *before* the constituents of a system have interacted is identical to the momenergy 4-vector figured *after* the constituents have interacted. In contrast, invariance of mass—the magnitude of the momenergy 4-vector—says that that mass is the same in *whatever* free-float frame it is figured

Momenergy: Is that a richer concept than mass?

Yes. Momenergy 4-vector reveals mass and more: the motion of object or system with the mass

Conservation of the momenergy of an isolated system: Does this imply that collisions and interactions within an isolated system cannot change the system's mass?

Yes. Mass of an isolated system, being the magnitude of its momenergy 4-vector, can never change (as long as the system remains isolated)

Conservation of the momenergy of an isolated system: Does this say that the constituents that enter a collision are necessarily the same in individual mass and in number as the constituents that leave that collision?

No! The constituents often change in a high-speed encounter.

**Example 1**: Collision of two balls of putty that stick together — after collision hotter and therefore very slightly more massive than before.

**Example 2**: Collision of two electrons (e-) with sufficient violence to create additional mass, a pair consisting of one ordinary electron and one positive electron (positron: e+):

$$e^- \text{ (fast)} + e^- \text{ (at rest)} \rightarrow e^+ + 3e^-$$

**Example 3**: Collision that radiates one or more photons:

$$e^- \text{ (fast)} + e^- \text{ (at test)} \rightarrow 2 \begin{pmatrix} \text{elections of} \\ \text{intetmediate} \\ \text{speed} \end{pmatrix} + 2 \begin{pmatrix} \text{electromagnetic} \\ \text{energy (photons)} \\ \text{emitted in the} \\ \text{collision ptocess} \end{pmatrix}$$

In all three examples the system momenergy and system mass are each the same before as after.

Can I figure the mass of an isolated system composed of a number, $n$, of freely-moving objects by simply adding the masses of the individual objects? **Example**: Collection of fast-moving molecules.

Ordinarily NO, but yes in one very special case: Two noninteracting objects move freely and in step, side by side. Then the mass of the system *does* equal the sum of the two individual masses. In the general case, where the system parts ate moving relative to each other, the relation between system mass and mass of parts is not additive. The length, in the sense of interval, of the 4-vector of total momenergy is not equal to the sum of the lengths of the individual momenergy 4-vectors, and for a simple reason: In the general case those vectors do not point in the same spacetime direction. *Energy* however, does add and momentum does add:

$$E_\text{system} = \sum_{i=1}^{n} E_i \quad \text{and} \quad p_{x,\text{system}} = \sum_{i=1}^{n} p_{x,i}$$

From these sums the mass of the system can be calculated

$$M_\text{system}^2 = E_\text{system}^2 - p_{x,\text{system}}^2 - p_{y,\text{system}}^2 - p_{z,\text{system}}^2$$

Can we simplify this expression for the mass of an isolated system composed of freely moving objects when we observe it from a free-float frame so chosen as to make the total momentum be zero?

Yes. In this case the mass of the system has a value given by the sum of energies of individual particles:

$$M_\text{system} = E_\text{system} = \sum_{i=1}^{n} E_i \quad \text{[in zero-total momentum frame]}$$

Moreover, the energy of each particle can always be expressed as sum of rest energy m plus kinetic energy K:

$$E_i = m + K_i, \qquad ( \; i \; = 1, \; 2, \; 3, \qquad , \; n \; )$$

So the mass of the system exceeds the sum of the masses of its individual particles by an amount equal to the total kinetic energy of all particles (but only as observed in the frame in which *total* momentum equals zero):

$$M_{system} = \sum_{i=1}^{n} m_i + \sum_{i=1}^{n} K_i$$

[in zero-total momentum frame]

For slow particles (Newtonian low-velocity limit) the kinetic energy term is negligible compared to the mass term. So it is natural that for years many thought that the mass of a system is the sum of the masses of its parts. However, such a belief leads to incorrect results at high velocities and is wrong as a matter of principle ar all velocities.

**What's the meaning of mass for a system in which the particles interact as well as move?**

The energies of interaction have to be taken into account. They therefore contribute to the total energy, $E_{system}$ that gives the mass

$$M_{\text{system}} = \left( E_{\text{system}}^2 - p_{\text{system}}^2 \right)^{1/2}$$

**How do we find out the mass of a system of particles (Table 8-1) that are held — or stick — together?**

Weigh it! Weigh it by conventional means if we are here on Earth and the system is small enough, otherwise by determining its gravitational pull on a satellite in free-float orbit about it.

**Does mass measure "amount of matter"?**

Nature does not offer us any such concept as "amount of matter." History has struck down every proposal to define such a term. Even if we could count number of atoms or by any other counting method try to evaluate amount of matter, that number would not equal mass. First, mass of the specimen changes with its temperature. Second, atoms tightly bonded in a solid weigh less — are less massive — than the same atoms free. Third, many of nature's atoms undergo radioactive decay, with still greater changes of mass. Moreover, around us occasionally, and conrinually in stars, the number of atoms and number of particles themselves undergo change. How then speak honestly? Mass, yes; "amount of matter," no.

**Does the explosion in space of a 20-mcgaton hydrogen bomb convert 0.93 kilogram of mass into energy (fusion, Section 8.7)?** [ $\Delta m = \Delta E_{\text{rest. conv.}}/c^2$ = (20 X $10^6$ tons TNT) X ($10^6$ grams/ton) X ($10^3$ calories/gram of "TNT equivalent") X (4.18 joules/calorie)A$^2$ = (8.36 X 1 $10^{16}$ joules)/(9 X $10^{16}$ meters$^2$/second$^2$) = 0.93 kilogram]

Yes *and* no! The question needs to be stated more carefully. Mass of the system of expanding gases, fragments, and radiation has the *same* value immediately after explosion as before; mass Al of the system has not changed. However, hydrogen has been transmuted to helium and other nuclear transformations have taken place. In consequence the *makeup* of mass of the system

$$M_{\text{system}} = \sum_{i=1}^{n} m_i + \sum_{i=1}^{n} K_i \qquad \begin{array}{l}\text{[in zero-total}\\ \text{momentum frame]}\end{array}$$

has changed. The first term on the right — sum of masses of individual constituents — has decreased by 0.93 kilogram:

$$\left(\sum_{i=1}^{n} m_i\right)_{\text{after}} = \left(\sum_{i=1}^{n} m_i\right)_{\text{before}} - 0.9.3 \text{ kilogram}$$

The second term — sum of kinetic energies, including "kinetic energy" of photons and neutrinos produced — has increased by the same amount:

$$\left(\sum_{i=1}^{n} m_i\right)_{\text{after}} = \left(\sum_{i=1}^{n} m_i\right)_{\text{before}} + 0.9.3 \text{ kilogram}$$

The first term on the right side of this equation — the original heat content of the bomb — is practically zero by comparison with 0.93 kilogram.

Thus part of the mass of constituents has been converted into energy; but the mass of the system has not changed.

The mass of the products of a nuclear fission explosion (Section 8.7: fragments of split nuclei of uranium, for example) — contained in an underground cavity, allowed to cool, collected, and weighed — is this mass less than the mass of the original nuclear device?

Yes! The key point is the waiting period, which allows heat and radiation to flow away until transmuted materials have practically the same heat content as that of original bomb. In the expression for the mass of the system

$$M_{\text{system}} = \sum_{i=1}^{n} m_i + \sum_{i=1}^{n} K_i \qquad \text{[in zero-total momentum frame]}$$

the second term on the right, the kinetic energy of thermal agitation — whose value rose suddenly at the time of explosion but dropped during the cooling period — has undergone no net alteration as a consequence of the explosion followed by cooling.

In contrast, the sum of masses

$$\sum m_i$$

has undergone a permanent decrease, and with it the mass $M$ of what one weighs (after the cooling period) has dropped (see the figure)

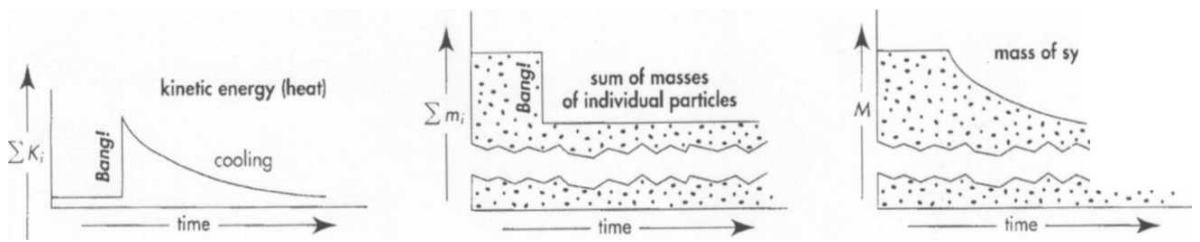

Does Einstein's statement that mass and energy are equivalent mean that energy is the same as mass?

No. Value of energy depends on the free-float frame of reference from which the particle (or isolated system of particles) is regarded. In contrast, value of mass is independent or inertial frame. Energy is only the *time* component of a momenergy 4-vector, whereas mass measures *entire magnitude* of that 4-vector. The time component gives the magnitude of the momenergy 4-vector only in the special case in which that 4-vector has no space component; that is, in a frame in which the momentum or the particle (or the total momentum of an isolated system of particles) equals zero. Only as measured in this special zero-momentum frame does energy have the same value as mass

Then what is the meaning of Einstein's statement that mass and energy are equivalent?

Einstein's statement refers to the reference frame in which the particle is at rest, so that it has zero momentum $p$ and zero kinetic energy $K$. Then $E = m + K \rightarrow m + 0$. In that case the energy is called the rest energy of the particle:

$$E_{rest} = m$$

In this expression, recall, the energy is measured in units of mass, for example kilograms. Multiply by the conversion factor $c^2$ to express energy in conventional units, for example joules (Table 7-1). The result is Einstein's famous equation:

$$E_{rest.conv} = mc^2$$

Many treatments of relativity fail to use the subscript "rest"' — needed to remind us that this equivalence of mass and energy refers only to the *rest* energy of the particle (for a system, the total energy in the zero-total-momentum frame).

Without delving into all fine points of legalistic phraseology, how significant is the conversion $c^2$ factor in the equation $E_{rest.conv} = mc^2$ '?

The conversion factor $c^2$, like the factor of conversion from seconds to meters or miles to feet (Box 3-2), today counts as a detail of convention, rather than as a deep new principle.

If the factor $c^2$ is not the central feature of the relationship between mass and energy, what *is* central?

The distinction between mass and energy is this: Mass is the magnitude of the momenergy 4-vector and energy is the time component of the same 4-vector. Any feature of any discussion that emphasizes this contrast is an aid to understanding. Any slurring of terminology that obscures this distinction is a potential source of error or confusion.

Is the mass of a moving object greater than the mass of the same object at rest?

No. It is the same whether the object is at rest or in motion; the same in all frames.

Really? Isn't the mass, $M$, of a system of freely moving particles given, not by the sum of the masses $m_i$ of the individual constituents, but by the sum of

Ouch! The concept of "relativistic mass" is subject to misunderstanding. Thar's why we don't use it. First, it applies the name mass — belonging to the energies $E$, *(but only in a frame in which total momentum of the system equals zero)*? Then why not give $E$, a new name and call it "relativistic mass" of the individual particle? Why not adopt the notation

$$m_{i,\,\text{rel}} = E_i = m_i + K_i \quad ?$$

With this notation, can't one then write

$$M = \sum_{i=1}^{n} m_{i,\,\text{rel}} \quad ? \text{ [in zero-total momentum frame]}$$

magnitude of a 4-vector — to a very different concept, the time component of a 4-vector. Second, it makes increase of energy of an object with velocity or momentum appear to be connected with some change in internal structure of the object. In reality, the increase of energy with velocity originates not in the object but in the geometric properties of space-time itself.

In order to make this point clear, should we call invariant mass of a particle its "rest mass"?

That is what we called it in the first edition of this book. But a thoughtful student pointed out that the phrase "rest mass" is also subject to misunderstanding: What happens to the "rest mass" of a particle when the particle **moves?** In reality mass is mass is mass. Mass has the same value in all frames, is invariant, no matter how the particle moves. [Galileo: "In questions of science **the** authority of a **Thou**sand is not worth the humble reasoning of a single individual."]

Can any simple diagram illustrate this contrast between mass and energy?

Yes. The figure shows the momentum-energy 4-vector of the same particle as measured in three different frames. Energy differs from frame to frame. Momentum differs from frame to frame. Mass (magnitude of 4-vector, represented by the length of handles on the arrows) has the same value, m = 8, in all frames.

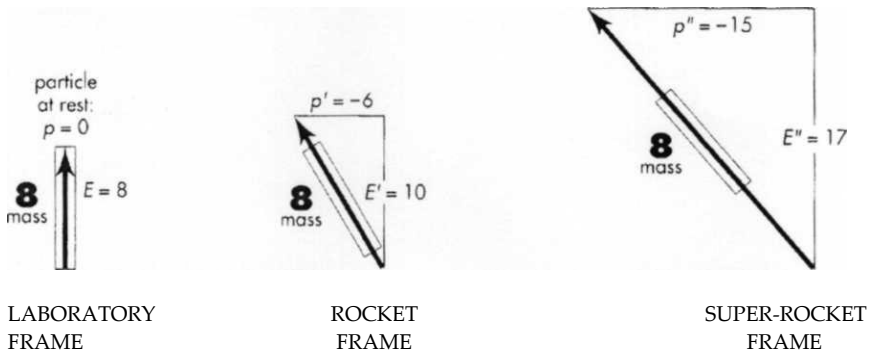

LABORATORY FRAME       ROCKET FRAME       SUPER-ROCKET FRAME

# ACKNOWLEDGMENTS

We thank colleagues old and young for the comments that helped us clarify, formulate, and describe the concept of mass in this chapter and in the final dialog, and very specially Academician Lev B. Okun, Institute of Theoretical and Experimental Physics, Moscow, for correspondence and personal discussions. We believe that our approach agrees with that in two of his articles, both entitled "The Concept of Mass," which appeared in *Physics Today*, June 1989, pages **31**-36, and *Soviet Physics-Uspekhi*, Volume 32, pages 629-638 (July 1989).